\documentclass[reqno]{amsart}

\usepackage{amsmath,amsfonts}
\usepackage{epsfig}
\usepackage{graphicx}

\def\alphaset{{\mathfrak A}}
\def\Av{{\rm Av}}
\def\cS{{\mathcal R}}
\def\duh{{\rm Duh}}
\def\Gspace{{\mathfrak G}}
 
\def\mus{{\mu_s}}

\def\opDelta{\widehat{\Delta}}
\def\opB{\widehat{B}}

\def\umu{{\underline{\mu}_s}}

\def\tr{{\rm Tr}}

\def\bra{\big\langle}
\def\ket{\big\rangle}
\def\Bra{\Big\langle}
\def\Ket{\Big\rangle}

\def\N{{\mathbb N}}

\def\R{{\mathbb R}}
\def\Z{{\mathbb Z}}

\def\uq{{\underline{q}}}

\def\uu{{\underline{u}}}

\def\ux{{\underline{x}}}

\def\cH{{\mathcal H}}

\def\cL{{\mathcal L}}
\def\cN{{\mathcal N}}
\def\cP{{\mathcal P}}

\def\Wspace{{\mathfrak{W}_{\xi_2}^\alpha(I)}}
\def\Wspacexi{{\mathfrak{W}_{\xi}^\alpha(I)}}
\def\Wspaceglob{{\widetilde{\mathfrak{W}}_{\xi_2,T_0}^1}}
\def\W1space{{\mathfrak{W}_{\xi_2}^1(I)}}

\def\1{{\bf 1}}

\def\eqnn{\begin{eqnarray*}}
\def\eeqnn{\end{eqnarray*}}
\def\eqn{\begin{eqnarray}}
\def\eeqn{\end{eqnarray}}

\def\prf{\begin{proof}}
\def\endprf{\end{proof}}

\theoremstyle{plain}
\newtheorem{theorem}{Theorem}[section]
\newtheorem{definition}[theorem]{Definition}
\newtheorem{proposition}[theorem]{Proposition}

\newtheorem{remark}[theorem]{Remark}

\numberwithin{equation}{section}

\begin{document}

\parskip=8pt

\title[Cauchy problem for the GP hierarchy]
{On the Cauchy problem for focusing and defocusing Gross-Pitaevskii hierarchies}
\author[T. Chen]{Thomas Chen}
\address{T. Chen,  
Department of Mathematics, University of Texas at Austin.}
\email{tc@math.utexas.edu}

\author[N. Pavlovi\'{c}]{Nata\v{s}a Pavlovi\'{c}}
\address{N. Pavlovi\'{c},  
Department of Mathematics, University of Texas at Austin.}
\email{natasa@math.utexas.edu}


\begin{abstract}
We consider the 
dynamical Gross-Pitaevskii (GP) hierarchy on $\R^d$, $d\geq1$,
for cubic, quintic, focusing and defocusing interactions. 
For both the focusing and defocusing case, and any $d\geq1$, 
we prove local
existence and uniqueness of solutions in certain 
Sobolev type spaces $\cH_\xi^\alpha$ of sequences of marginal 
density matrices.
The regularity is accounted for by  
$$
	\alpha \,  \left\{
	\begin{array}{rcl}
	> &\frac12& {\rm if} \; d=1 \\ 
	>&\frac d2-\frac{1}{2(p-1)} & {\rm if} \; d\geq2 \; {\rm and} \; (d,p)\neq(3,2)\\
	\geq & 1 & {\rm if} \; (d,p)=(3,2) \,,
	\end{array}
	\right.
$$
where $p=2$ for the cubic, and $p=4$ for the quintic GP hierarchy;
the parameter $\xi>0$ is arbitrary and determines the energy scale of the problem.
This result includes the proof of an a priori spacetime bound conjectured
by Klainerman and Machedon for the cubic GP hierarchy in $d=3$.
In the defocusing case, we prove the existence and uniqueness of solutions
globally in time for the cubic GP hierarchy for $1\leq d\leq3$,
and of the quintic GP hierarchy for $1\leq d\leq 2$,
in an appropriate space of Sobolev type, and under the assumption of an a priori
energy bound.
For the focusing GP hierarchies, we prove lower bounds on the blowup rate.
Also pseudoconformal invariance is established in the cases corresponding to $L^2$ criticality,
both in the focusing and defocusing context. 
All of these results hold without the assumption of factorized initial conditions.
\end{abstract}

\maketitle

\section{Introduction}

The derivation of the nonlinear Schr\"odinger equation as the dynamical mean
field limit of the manybody quantum dynamics of interacting Bose gases
is a research area that is recently experiencing
remarkable progress, see \cite{esy1,esy2,ey,kiscst,klma,rosc} and the references therein,
and also \cite{adgote,eesy,frgrsc,frknpi,frknsc,he,sp}.
A main motivation to investigate this problem is to understand the dynamical behavior of
Bose-Einstein condensates. 
For recent developments in the mathematical analysis of
Bose gases and their
condensation, we refer to the fundamental work of 
Lieb, Seiringer, Yngvason, et al.; see
\cite{ailisesoyn,lise,lisesoyn,liseyn} and the references therein.

The  procedure  developed in the landmark works of Erd\"os, Schlein, and Yau, 
\cite{esy1,esy2,ey}, to obtain the dynamical mean field limit of an
interacting Bose gas, comprises the following main ingredients.
One determines the BBGKY hierarchy of 
marginal density matrices
for particle number $N$, and derives the Gross-Pitaevskii (GP)
hierarchy in the limit $N\rightarrow\infty$, for a scaling where
the particle interaction potential tends to a delta distribution; see also \cite{kiscst,sc}.
For factorized initial data, the
solutions of the GP hierarchy are governed by a cubic NLS for
systems with 2-body interactions, \cite{esy1,esy2,ey,kiscst}, 
and quintic NLS for systems with 3-body interactions, \cite{chpa}.
The proof of the uniqueness of solutions of the GP hierarchy 
is the most difficult part of this analysis, and is obtained in \cite{esy1,esy2,ey}
by use of highly sophisticated Feynman graph expansion methods inspired by quantum field theory.

Recently, an alternative method to prove the uniqueness of solutions in the $d=3$ case
has been developed by Klainerman and Machedon in \cite{klma}, using spacetime bounds on the density
matrices in the GP hierarchy; this result makes the assumption of a particular a priori 
spacetime bound on the density matrices which has so far remained conjectural.  
In the work \cite{kiscst} of Kirkpatrick, Schlein, and Staffilani,
the corresponding problem in $d=2$ is solved,
and the assumption made in \cite{klma} is replaced by a spatial a
priori bound which is proven in \cite{kiscst}.  
Alternative methods to obtain dynamical mean field limits of interacting
Bose gases using operator-theoretic methods are developed by Fr\"ohlich et al in \cite{frgrsc,frknpi,frknsc}.

All of the above mentioned works discuss Bose gases with {\em repulsive}
interactions; it is currently not known how to obtain a GP hierarchy from the
$N\rightarrow\infty$ limit of a BBGKY hierarchy with attractive
interactions. 
In the work at hand, 
we have nothing to add to this issue.
Instead, we start here directly from the level of the GP hierarchy,
and are thus free to also consider {\em attractive} interactions within this context.
Accordingly, we will refer to the corresponding GP hierarchies as   {\em cubic},
{\em quintic}, {\em focusing}, or {\em defocusing GP hierarchies}, 
depending on the type of the NLS governing the solutions obtained from
factorized initial conditions.


In the present work, we investigate the Cauchy problem for the cubic and quintic GP hierarchy
with focusing and defocusing interactions. Our results do not assume
any factorization of the initial data.
As a crucial ingredient of our arguments, we introduce Banach spaces 
$\cH_\xi^\alpha=\{ \, \Gamma\in\Gspace \,  | \, \| \, \Gamma \, \|_{\cH_\xi^\alpha} <\infty \, \}$  
where
\eqn\label{bigG}
	\Gspace \, = \, \{ \, \Gamma \, = \, ( \, \gamma^{(k)}(x_1,\dots,x_k;x_1',\dots,x_k') \, )_{k\in\N} 
	\, | \,
	\tr \gamma^{(k)} \, < \, \infty \, \}
\eeqn
is the space of sequences of $k$-particle density matrices, and
\eqn
	\| \, \Gamma \, \|_{\cH_\xi^\alpha} \, := \,
	\sum_{k\in\N} \xi^k \, \| \, \gamma^{(k)} \, \|_{H^\alpha(\R^{dk}\times\R^{dk})} \,.
\eeqn
The parameter $\xi>0$ is determined by the initial condition, and it sets the energy scale of 
a given Cauchy problem.
If $\Gamma\in\cH_\xi^\alpha$, then $\xi^{-1}$ is the typical $H^\alpha$-energy per particle.

The parameter $\alpha$ determines the regularity of the solution,
and our results hold for  $\alpha\in\alphaset(d,p)$ where
\eqn\label{eq-alphaset-def-0}
	\alphaset(d,p) \, := \, \left\{
	\begin{array}{cc}
	(\frac12,\infty) & {\rm if} \; d=1 \\ 
	(\frac d2-\frac{1}{2(p-1)}, \infty) & {\rm if} \; d\geq2 \; {\rm and} \; (d,p)\neq(3,2)\\
	\big[1,\infty) & {\rm if} \; (d,p)=(3,2) \,,
	\end{array}
	\right.
\eeqn 
in dimensions $d\geq1$, and where $p=2$ for the cubic, and $p=4$ for the quintic GP hierarchy.
The parameter $\xi>0$ determines the energy scale of the problem.

The main results proven in this paper are:
\begin{enumerate}
\item
We prove local existence and uniqueness of solutions
for the cubic and
quintic GP hierarchy with focusing or defocusing interactions, 
in $\cH_\xi^\alpha$, for $\alpha\in \alphaset(d,p)$,
which satisfy a spacetime bound $\|\opB\Gamma\|_{L^1_{t\in I}\cH^{\alpha}_{\xi}}<\infty$
for some $\xi>0$ (the operator $\opB$ is defined in Section \ref{sec-defandresults-1}
below). This spacetime bound  has been
conjectured by Klainerman and Machedon in \cite{klma}.
It is of Strichartz-type, and is proven in Section \ref{sec-locwp-1} 
using a Picard-type fixed point argument on the space $L^1_{t\in [0,T]}\cH_\xi^\alpha$; 
see inequality
(\ref{eqn-BGamma-spacetime-bd-0}) and Remark \ref{rem-Strichartz-1} below. 

Accordingly, we conclude that a solution of the GP hierarchy in $\cH_\xi^\alpha$
is unique if and only if
this spacetime bound holds. 
\\

\item
We prove the global existence and uniqueness of solutions
in $\cH_\xi^1$ satisfying the above noted spacetime bound,
for the defocusing cubic GP hierarchy for $1\leq d\leq3$,
and the defocusing quintic GP hierarchy for $1\leq d\leq 2$, provided that an
a priori bound $\|\Gamma(t)\|_{\cH_\xi^1}<c$ holds for $\xi>0$
sufficiently small.
\\

\item
We indroduce generalized pseudoconformal transformations, and prove the 
invariance
of the cubic GP hierarchy in $d=2$, and of the quintic GP hierarchy in $d=1$,
under their application. Because the NLS obtained from factorized initial data
in these cases are $L^2$-critical, we will, for brevity, refer to these GP hierarchies as being
$L^2$-critical.
\\

\item
For the focusing cubic or quintic GP hierarchy, 
we prove lower bounds on the blowup rate in $\cH^\alpha_\xi$ and $\cL^r_\xi$,
where both spaces are defined in Section \ref{sec-defandresults-1} below.
\end{enumerate}

An important ingredient of our proof of the local existence and uniqueness of
solutions  is the use of certain spacetime bounds for the non-interacting GP hierarchy established
in \cite{klma} for the cubic GP hierarchy in $d=3$ (which were generalized 
to cubic in $d=2$ in \cite{kiscst}, and to the quintic GP hierarchy in \cite{chpa}), 
and the ``boardgame estimates" developed in \cite{klma} (and generalized to the quintic case in \cite{chpa}),
which were motivated by the Feynman graph expansion techniques of \cite{esy1,esy2}. 
For our discussion of blowup solutions of the focusing (cubic or quintic) GP hierarchy,
we make extensive use of a quantity that controls the average $H^\alpha$-energy per particle, 
and, in a different form, the
average $L^r$-norm per particle. It is introduced in Definition \ref{def-AvHLp-1} below,
and turns out to be the key observable for our discussion of blowup solutions.  
\\

\subsection*{Organization of the paper}

In Section \ref{sec-defandresults-1}, we introduce the cubic and quintic GP hierarchy,
and state our main theorems.
In Section \ref{sec-locwp-1}, we prove the local wellposedness of the Cauchy problem
for the cubic and quintic GP hierarchy, for both focusing and defocusing interactions.
In Section \ref{sec-gwp-GP-1}, the local wellposedness is enhanced to global wellposedness
for the cubic and quintic defocusing GP hierarchies, using energy conservation.
In Section \ref{sec-Proof-blowuprate-1}, we prove lower bounds on the blowup rate
of blowup solutions in the spaces $\cH_\xi^\alpha$ and $\cL^r_\xi$ (see below for their
definitions).
In Section \ref{sec-confinv-1}, we prove the pseudoconformal invariance of the $L^2$-critical
cubic (in $d=2$) and quintic (in $d=1$) GP hierarchies.
In the Appendix, we reformulate the Klainerman-Machedon spacetime bounds in a form
convenient for our work.
 
\newpage

\section{Definition of the model and statement of the main results}
\label{sec-defandresults-1}

We introduce the space 
\eqn
	\Gspace \, := \, \bigoplus_{k=1}^\infty L^2(\R^{dk}\times\R^{dk})  
\eeqn
of sequences of density matrices
\eqn
	\Gamma \, := \, (\, \gamma^{(k)} \, )_{k\in\N}
\eeqn
where $\gamma^{(k)}\geq0$, $\tr\gamma^{(k)} =1$,
and where every $\gamma^{(k)}(\ux_k,\ux_k')$ is symmetric in all components of $\ux_k$,
and in all components of $\ux_k'$, respectively.

We call $\Gamma=(\gamma^{(k)})_{k\in\N}$  admissible if   
\eqn
	\lefteqn{
	\gamma^{(k)}(\ux_k;\ux_k') 
	}
	\\
	&&\, = \, \int dx_{k+1} \cdots dx_{k+\frac p2}
	\, \gamma^{(k+\frac{p}{2})}(\ux_{k},x_{k+1},\dots,x_{k+\frac p2};\ux_k',x_{k+1},\dots,x_{k+\frac p2}) 
	\nonumber
\eeqn  
for all $k\in\N$.

Let $0<\xi<1$ and $r>1$.
We define
\eqn
	\cL_\xi^r \, := \, \Big\{ \, \Gamma \, \in \, \Gspace \, \Big| \, \|\Gamma\|_{\cL_\xi^r} < \, \infty \, \Big\}
\eeqn
where
\eqn
	\|\Gamma\|_{\cL_\xi^r} \, := \, \sum_{k=1}^\infty \xi^{ k} 
	\| \, \gamma^{(k)} \, \|_{L^r(\R^{dk}\times\R^{dk})} \,.
\eeqn 
Furthermore, we define
\eqn
	\cH_\xi^\alpha \, := \, \Big\{ \, \Gamma \, \in \, \Gspace \, \Big| \, \|\Gamma\|_{\cH_\xi^\alpha} < \, \infty \, \Big\}
\eeqn
where
\eqn
	\|\Gamma\|_{\cH_\xi^\alpha} \, = \, \sum_{k=1}^\infty \xi^{ k} 
	\| \,  \gamma^{(k)} \, \|_{H^\alpha(\R^{dk}\times\R^{dk})} \,,
\eeqn
with
\eqn
	\| \, \gamma^{(k)} \, \|_{H^\alpha(\R^{dk}\times\R^{dk})} \, = \,
	\| \, S^{(k,\alpha)} \, \gamma^{(k)} \, \|_{L^2(\R^{dk}\times\R^{dk})}  \,,
\eeqn
and $S^{(k,\alpha)}:=\prod_{j=1}^k\langle\nabla_{x_j}\rangle^\alpha\langle\nabla_{x_j'}\rangle^\alpha$.
 
Clearly, $\cL_\xi^r$,  $\cH_\xi^\alpha$ are Banach spaces.
 
We note that Banach spaces of integral kernels of a similar type as those introduced
above are, for instance, 
used for operator-theoretic renormalization group methods 
in the spectral analysis of quantum electrodynamics,
\cite{bcfs}.

Let $p\in\{2,4\}$.  
We consider the $p$-GP (Gross-Pitaevskii) hierarchy given by
\eqn \label{eq-def-b0-2}
	i\partial_t \gamma^{(k)} \, = \, \sum_{j=1}^k [-\Delta_{x_j},\gamma^{(k)}]   
	\, + \,  \mu B_{k+\frac p2} \gamma^{(k+\frac p2)}
\eeqn
in $d$ dimensions, for $k\in\N$. Here,
\eqn
	\lefteqn{
	\left(B_{k+\frac p2}\gamma^{(k+\frac{p}{2})}\right)(t,x_1,\dots,x_k;x_1',\dots,x_k')
	}
	\\
	&& := \,
	\sum_{j=1}^k \left(B_{j;k+1,\dots,k+\frac p2}\gamma^{(k+\frac{p}{2})}\right)(t,x_1,\dots,x_k;x_1',\dots,x_k')
	\nonumber\\
	&& := \, \sum_{j=1}^k\int dx_{k+1}\cdots dx_{k+\frac p2} dx_{k+1}'\cdots dx_{k+\frac p2}' \,
	\nonumber\\
	&&\quad\quad\quad\quad 
	\big[ \, \prod_{\ell=k+1}^{k+\frac p2} \delta(x_j-x_{\ell})\cdots \delta(x_j-x_{\ell}' )  
	- \prod_{\ell=k+1}^{k+\frac p2}\delta(x_j'-x_{\ell})\cdots\delta(x_j'-x_{\ell}') \, \big] 
	\nonumber\\
	&& \quad\quad\quad\quad\quad\quad 
	\gamma^{(k+\frac p2)}(t,x_1,\dots,x_{k+\frac p2};x_1',\dots,x_{k+\frac p2}') \, 
	\nonumber\, 
\eeqn 
accounts for  $\frac p2+1$-body interactions between the Bose particles.

For a factorized initial condition
\eqn\label{eq-initcond-fact-1}
	\gamma^{(k)}(0) \, = \, | \, \phi_0 \, \rangle \langle \, \phi_0 \, |^{\otimes k}
\eeqn
with $\phi_0\in H^\alpha$, one obtains that
\eqn
	\gamma^{(k)}(t) \, = \, | \, \phi(t) \, \rangle \langle \, \phi(t) \, |^{\otimes k}
\eeqn
is a solution of \eqref{eq-def-b0-2} if $\phi_t$ satisfies the NLS
\eqn\label{eq-NLS-p-1}
	i\partial_t \phi_t \, + \, \Delta_x \phi_t \, - \, \mu \, |\phi_t|^p \, \phi_t \, = \, 0
\eeqn
with initial condition $\phi(0)=\phi_0$, where $\mu\in\{1,-1\}$.
For $p=2$, this is the cubic NLS, and for $p=4$, this is the quintic NLS.
The NLS is defocusing for $\mu=1$, and focusing for $\mu=-1$. 

Accordingly,
we refer to (\ref{eq-def-b0-2}) as the {\em cubic GP hierarchy} if $p=2$,
and as the {\em quintic GP hierarchy} if $p=4$.
Moreover, for $\mu=1$ or $\mu=-1$ we refer to the GP hierarchies as being  
defocusing or focusing, respectively.

We recall the definition of the set $\alphaset(d,p)$, for $p=2,4$ and $d\geq1$,  
\eqn\label{eq-alphaset-def-1}
	\alphaset(d,p) \, = \, \left\{
	\begin{array}{cc}
	(\frac12,\infty) & {\rm if} \; d=1 \\ 
	(\frac d2-\frac{1}{2(p-1)}, \infty) & {\rm if} \; d\geq2 \; {\rm and} \; (d,p)\neq(3,2)\\
	\big[1,\infty) & {\rm if} \; (d,p)=(3,2)
	\end{array}
	\right.
\eeqn 
Our main result in this paper is the following theorem.

\begin{theorem}\label{thm-main-0}
Let $0<\xi_2=\eta\xi_1\leq\xi_1<1$.  
Assume that $\alpha\in\alphaset(d,p)$ where $d\geq1$ and $p\in\{2,4\}$, and $0<\eta<1$ sufficiently small.
Then, the following hold.
\begin{itemize}
\item[(i)]
For every $\Gamma_0\in\cH_{{\xi_1}}^\alpha$, there
exist constants $T>0$ and $0<\xi_2\leq\xi_1$ 
such that there exists a unique
solution $\Gamma(t)\in\cH_{\xi_2}^\alpha$ for $t\in[0,T]$ with 
$\|\opB\Gamma\|_{L^1_{t\in[0,T]}\cH_{\xi_2}^\alpha}<\infty$.  
\\
\item[(ii)]
Assume that given $\Gamma_0\in\cH_{{\xi_1}}^\alpha$, there are constants $T>0$ 
and $0<\xi_2\leq\xi_1$ such that for $t\in I=[0,T]$, there exists a  
solution $\Gamma(t)$ of the $p$-GP hierarchy (\ref{seclwp-pGP})
in the space $L^\infty_{t\in I}\cH_{\xi_2}^\alpha$. 
\\
\\
Then, the solution $\Gamma(t)\in L^\infty_{t\in I}\cH_{\xi_2}^\alpha$ 
is {\em unique} if and only if 
$\| \, \opB \Gamma \, \|_{L^1_{t\in I}\cH_{\xi}^\alpha}<\infty$ 
holds for some $\xi>0$.
\\
\\
If the latter is satisfied, then in fact, the Strichartz-type bound
\eqn\label{eqn-BGamma-spacetime-bd-0}
	\|\opB\Gamma\|_{L^1_{t\in I}\cH_{\xi_2}^\alpha}
	&\leq&C(d,p,\xi_1,\xi_2) \,
	\|\Gamma_0\|_{\cH_{\xi_1}^\alpha}  
\eeqn
holds.
\\
\end{itemize}
\end{theorem}

\begin{remark}\label{rem-BBGKYlim-1}
An immediate implication of part (ii) of Theorem \ref{thm-main-0} is that 
every solution $\Gamma(t)$ extracted by a diagonal argument 
from the $N\rightarrow\infty$ limit of the $N$-particle BBGKY hierarchies
(with repulsive interactions)
studied by Erd\"os-Schlein-Yau in \cite{esy1,esy2}, 
Kirkpatrick-Schlein-Staffilani in
\cite{kiscst}, and Chen-Pavlovi\'c in \cite{chpa}, satisfies 
$\| \, \opB \Gamma \, \|_{L^1_{t\in I}\cH_{\xi}^\alpha}<\infty$ 
for some $\xi>0$.
This is true because uniqueness of those solutions has been established in these
works with independent methods.
\end{remark}

 
\begin{remark}
The role of the parameters $\xi_1,\xi_2$ is as follows: 
Given initial data $\Gamma_0=(\gamma^{(k)})_{k\in\N}$ with 
$\|\gamma^{(k)}\|_{H^\alpha(\R^{dk}\times\R^{dk})}<\infty$ for all $k$, 
we determine $\xi_1>0$ sufficiently small such that
$\Gamma_0\in\cH_{\xi_1}^\alpha$.
This means that the energy per particle in $\Gamma_0$ is bounded by $\xi_1^{-1}$.
In cases of physical interest, $\xi_1>0$;
the notion of an energy per particle will be quantified below.
Then, we find a suitable $\xi_2=\eta\xi_1\ll\xi_1$ such that the
Cauchy problem for the  the GP hierarchy can be solved in a sufficiently large space
$\cH^\alpha_{\xi_2}$.
The requirement $\xi_2 \ll \xi_1$ is used to ensure
that a solution $\Gamma(t)$ does not drift out of  $\cH^\alpha_{\xi_2}$ for 
$t\in I=[0,T]$ with $T=T(\xi_2)>0$; we thereby impose the assumption that the energy per particle does not
exceed $\xi_2^{-1}$ while $t\in I$, but once this assumption is
violated, we may choose $0<\xi_2'<\xi_2$ to continue the solution to  $T(\xi_2')>T(\xi_2)$.
\end{remark}

\begin{remark}
In particular, there is no implication of the size of $\xi_2$ on the regularity
accounted for by $\alpha$.
For factorized initial data, the statement that the solution of the NLS remains in $H^\alpha$
for $t\in I$ is equivalent to the statement that the solution of the GP hierarchy
remains in $\cH^\alpha_{\xi}$ for {\em an arbitrary} nonzero $\xi>0$. 
\end{remark}

\begin{remark}\label{rem-Strichartz-1}
We note that the estimate (\ref{eqn-BGamma-spacetime-bd-0}), for the cubic GP hierarchy
with $d=3$ and $\alpha=1$, proves the a priori spacetime bound conjectured in \cite{klma}.
For factorized initial data $\Gamma=(|\phi_0\rangle\langle\phi_0|^{\otimes k})_{k\in\N}$ 
in the cubic case, so that $\Gamma=(|\phi(t)\rangle\langle\phi(t)|^{\otimes k})_{k\in\N}$
where $i\partial_t\phi+\Delta\phi-\mu|\phi|^2\phi=0$, it corresponds to the inequality
\eqn
	\| \, |\phi|^2\phi \, \|_{L^1_{t\in I}H^\alpha}^{\frac13} \, \leq \, C(T) \, \| \, \phi_0 \, \|_{H^\alpha}
\eeqn
which is of Strichartz type. 
The example of factorized solutions with $\phi(t)\in H^1$, $t\in I$, is discussed in
detail in \cite{klma}.
\end{remark}

\begin{definition}
We say that a solution $\Gamma(t)$ of the GP hierarchy blows up in finite time 
with respect to $H^\alpha$ if 
there exists $T^*<\infty$  such that for every $\xi>0$ there exists $T_{\xi,\Gamma}^*<T^*$ such that
$\|\Gamma(t)\|_{\cH_{\xi}^\alpha}\rightarrow\infty$ as $t\nearrow T^{*}_{\xi,\Gamma}$,
and $T_{\xi,\Gamma}^*\nearrow T^*$ as $\xi\rightarrow0$. 
\end{definition}

For the study of blowup solutions, it is convenient to introduce the following quantity.

\begin{definition}\label{def-AvHLp-1}
We refer to
\eqn
	\Av_{H^\alpha}(\Gamma ) \, := \, \Big[ \, \sup\big\{ \, \xi >0 \, \big| \,
	\| \, \Gamma \, \|_{\cH_\xi^\alpha} <\infty\, \big\} \, \Big]^{-1} \,,
\eeqn
\eqn
	\Av_{L^r}(\Gamma ) \, := \, \Big[ \, \sup\big\{ \, \xi >0 \, \big| \,
	\| \, \Gamma \, \|_{\cL_\xi^r} <\infty\, \big\} \, \Big]^{-1} \,,
\eeqn
respectively, as the typical (or average) $H^\alpha$-energy and the typical $L^r$-norm per particle. 
\end{definition}

We note that 
\eqn
	\Gamma \, = \, ( \, | \, \phi \, \rangle \langle \, \phi \, |^{\otimes k} \, )_{k\in\N}
	\; \; \; \Rightarrow \; \; \; 
 	\Av_{H^\alpha}(\Gamma ) \, = \, \|\phi\|_{H^\alpha}^2 \mbox{ and } 
	\Av_{L^r}(\Gamma ) \, = \, \|\phi\|_{L^r}^2
\eeqn
in the factorized case.

The fact that $\Gamma\in\cH_\xi^\alpha$ means that
the typical energy per particle is bounded by $\Av_{H^\alpha}(\Gamma )<\xi^{-1}$.
Therefore, the parameter $\xi$ determines the $H^\alpha$-energy scale in the problem.
While solutions with a bounded $H^\alpha$-energy remain in the same $\cH_\xi^\alpha$,
blowup solutions undergo transitions
$\cH_{\xi_1}^\alpha\rightarrow\cH_{\xi_2}^\alpha\rightarrow\cH_{\xi_3}^\alpha\rightarrow\cdots$
where the sequence $\xi_1>\xi_2>\cdots$ converges to zero as $t\rightarrow T^*$.

It is easy to see that blowup in finite time of $\Gamma(t)$ with respect to $H^\alpha$
is equivalent to the statement that $\Av_{H^\alpha}(\Gamma(t))\rightarrow\infty$
as $t\nearrow T^*$.

Clearly, $(\Av_{N}(\Gamma ))^{-1}$ is the convergence radius of $\|\Gamma\|_{\cN_\xi}$
as a power series in $\xi$, for the norms $N=H^\alpha,L^r$ and 
$\cN_\xi=\cH^\alpha_\xi ,\cL^r_\xi$, respectively.

\begin{theorem}
\label{thm-blowuprate-L2crit-1}
Assume that $\Gamma(t)$ is a solution of the (cubic $p=2$ or $p=4$ quintic) $p$-GP hierarchy
with initial condition $\Gamma(t_0)=\Gamma_{0}\in\cH_{\xi}^\alpha$,
for some $\xi>0$, which blows up in finite time. Then, the following
lower bounds on the blowup rate hold:
\begin{enumerate}
\item[($a$)] 
Assume that $\frac4d\leq p < \frac{4}{d-2\alpha}$. Then,
\eqn
	( \, \Av_{H^\alpha}(\Gamma(t)) \, )^{\frac12}  
	\, > \, \frac{C}{|T^*-t|^{(2\alpha-d+\frac4p)/4}} \,.
\eeqn 
Thus specifically, for the cubic GP hierarchy in $d=2$, and for the quintic GP hierarchy in $d=1$,
\eqn
	( \, \Av_{H^1}(\Gamma(t )) \, )^{\frac12} \, \geq \, \frac{C}{|t -T^*|^{\frac12}} \,,
\eeqn 
with respect to the Sobolev spaces $H^\alpha$, $\cH_\xi^\alpha$.

\item[($b$)] 
\eqn
	( \, \Av_{L^r}(\Gamma(t )) \, )^{\frac12} \, \geq \, 
        \frac{C}{|t -T^*|^{\frac{1}{p}-\frac{d}{2r}}} \,, \mbox{ for } \frac{pd}{2} < r.
\eeqn 
\end{enumerate} 
\end{theorem}

\begin{remark}
We note that in the factorized case, the above 
lower bounds on the blow-up rate coincide with the known
lower bounds on the blow-up rate for solutions to the NLS (see, for example, \cite{ca}).
\end{remark} 

The cubic GP hierarchy in $d=2$, and the quintic GP hierarchy in $d=1$
are distinguished by being invariant under a class of generalized 
pseudoconformal transformations, as presented below.
Let us first recall pseudoconformal invariance 
on the level of the NLS (\ref{eq-NLS-p-1}).
If the NLS (\ref{eq-NLS-p-1}) is $L^2$-critical, that is, $p=\frac 4d$,
it is invariant under the pseudoconformal transformations
\eqn\label{eq-NLSpc-def-1}
	{\mathcal P}\phi_t (x)
	\, := \, 
	\frac{1}{(1+bt)^{1/2}} \, e^{-i\frac{b x^2 }{1+bt}}
	\phi_{\frac{1}{1+bt}}\Big( \, \frac{x}{1+bt} \, \Big) \,,
\eeqn
for $b\in\R\setminus\{0\}$. 
That is,
\eqn
	i\partial_t{\mathcal P}\phi_t \, + \, \Delta {\mathcal P}\phi_t 
	\, - \, \mu \, |{\mathcal P}\phi_t|^p \, {\mathcal P}\phi_t \, = \, 0 \,;
\eeqn
see for instance \cite{ca}.
There are two cases of $L^2$-critical NLS with $p\in\N$: The cubic ($p=2$)
NLS in $d=2$, and the quintic ($p=4$) NLS in $d=1$.

For the GP hierarchy, one can likewise introduce pseudoconformal transformations,
and as we prove in this paper,
the GP hierarchy is pseudoconformally invariant when
$p=2$ and $d=2$ (cubic), or
$p=4$ and $d=1$ (quintic).
This property is independent of whether the GP hierarchy is defocusing, $\mu=1$,
or focusing, $\mu=-1$.

\begin{theorem}
\label{thm-pseudoconf-gamma-1}
For $d=2$ and $p=2$ (cubic),
or $d=1$ and $p=4$ (quintic),
the focusing or defocusing ($\mu\in\{1,-1\}$)
GP hierarchy \eqref{eq-def-b0-2} is invariant under the pseudoconformal 
transformations 
\eqn\label{eq-pseudoconf-gamma-1}
	\lefteqn{
	\cP\gamma^{(k)} (t,\ux_k;\ux_k')
	}
	\nonumber\\
	&&
	\, := \, 
	\frac{1}{(1+bt)^{dk}} \, e^{-i\frac{b(|\ux_k|^2-|\ux_k'|^2)}{1+bt}}
	\gamma^{(k)}\Big( \, \frac{1}{1+bt} \, , \, \frac{\ux_k}{1+bt} \, ; \, \frac{\ux_k'}{1+bt} \, \Big) \,,
\eeqn
for $b\in\R\setminus\{0\}$.

That is,
\eqn\label{eq-infhierarch-pseudoconf-1}
	i\partial_t \cP\gamma^{(k)} \, + \, 
	\Delta_\pm^{(k)}\cP\gamma^{(k)}    
	\, - \, \mu \,  B_{k+\frac p2} \cP\gamma^{(k+2)} = 0 \,,
\eeqn
for all $k\geq1$. 
\end{theorem}

The proof is given in Section \ref{sec-confinv-1}.
For a survey of related matters for the NLS, see for instance \cite{ca,ra,ta}.

Of course, the following is immediately clear.

\begin{theorem}
Assume that $\alpha\in\alphaset(d,p)$ where $d\geq1$ and $p\in\{2,4\}$.
Moreover, assume that $\Gamma(t)\in\cH_{\xi_2}^\alpha$ solves the (cubic or quintic)
focusing ($\mu=-1$) GP hierarchy with factorized
initial condition $\Gamma_0=(|\phi_0\rangle\langle\phi_0|^{\otimes k})_{k\in\N}\in\cH_{\xi }^\alpha$
for some $\xi>0$,
where $\phi_0\in H^\alpha $. 
\\
\\
Then, if there exists $T^*<\infty$
such that $\|\phi(t)\|_{H^\alpha}\rightarrow\infty$ as $t\nearrow T^*$,
it follows that also $\Av_{H^\alpha}(\Gamma(t))\rightarrow\infty$  as $t\nearrow T^*$.
\end{theorem}

\prf
This follows from $\Av_{H^\alpha}(\Gamma(t))=\|\phi(t)\|_{H^\alpha}^2$ for product states.
\endprf

For various scenarios in which blowup occurs for solutions of the cubic 
or quintic NLS, we refer to the literature; see for instance \cite{ca,ra} for
surveys.

\newpage

\section{Local existence and uniqueness of solutions for the focusing and defocusing GP hierarchy}
\label{sec-locwp-1}

In this section, we prove a local existence and uniqueness result for the cubic and quintic
GP hierarchy for both focusing and defocusing interactions. We formulate all arguments for the cubic hierarchy ($p=2$). 
For the quintic hierarchy ($p=4$), the generalizations are straightforward,
and will only be briefly described. 

We introduce the notation
\eqn
	\Delta_{\pm}^{(k)} \, = \, \Delta_{\ux_k} - \Delta_{\ux'_k}
\eeqn
with
\eqn	
	\Delta_{\ux_k} \, = \, \sum_{j=1}^k\Delta_{x_j}
\eeqn
and
\eqn
	\Delta_{\pm, x_j} \, = \, \Delta_{x_j} - \Delta_{x_j'}  \,.
\eeqn
Moreover, we write
\eqn
	\opDelta_\pm \Gamma \, := \, ( \, \Delta^{(k)}_\pm \gamma^{(k)} \, )_{k\in\N}
\eeqn
and 
\eqn
	\opB \Gamma \, := \, ( \, B_{k+1} \gamma^{(k+1)} \, )_{k\in\N} \,.
\eeqn
Then, the $p$-GP hierarchy \eqref{eq-def-b0-2} can be written as   
\eqn \label{seclwp-pGP}
        i\partial_t \Gamma \, + \, \opDelta_\pm \Gamma \, = \, \mu \opB \Gamma \,.
\eeqn 
In integral formulation, it is formally given by
\eqn \label{int-pGP} 
	\Gamma(t) \, = \, e^{it\opDelta_\pm}\Gamma_0 \, - \, i \mu 
        \int_0^t ds \, e^{i(t-s)\opDelta_\pm} \opB\Gamma(s) \,. 
\eeqn
We will prove the local existence and uniqueness of such a solution in the following manner.
We note that (\ref{int-pGP}) can be formally written as a system of integral equations
\eqn  
		\Gamma(t) & = & e^{it\opDelta_\pm}\Gamma_0 \, - \, i \mu 
        \int_0^t ds \, e^{i(t-s)\opDelta_\pm} \opB\Gamma(s)
        \label{int-pGP-sys-1}\\
        \opB\Gamma(t) & = & \opB \, e^{it\opDelta_\pm}\Gamma_0 \, - \, i \mu 
        \int_0^t ds \, \opB \, e^{i(t-s)\opDelta_\pm} \opB\Gamma(s) \,, 
        \label{int-pGP-sys-2}
\eeqn
where we note that the second line is formally a self-consistent fixed point equation for $\opB\Gamma$.

Let $I:=[0,T]$. We introduce the product space
\eqn\label{def-Wspace-1}
	\Wspacexi \, := \,  
	L^\infty_{t\in I}\cH_\xi^\alpha \, \times \,
	L^1_{t\in I}\cH_\xi^\alpha \,  ,
\eeqn
which we endow with the norm
\eqn
	\| \, (\Gamma, \Xi) \, \|_{\Wspacexi} \, := \, 
	\| \, \Gamma \, \|_{L^\infty_{t\in I}\cH_\xi^\alpha}
	\, + \, 
	\| \, \Xi \, \|_{L^1_{t\in I}\cH_\xi^\alpha} \,.
\eeqn
Clearly, $( \Wspacexi , \| \, \cdot \, \|_{\Wspacexi} )$ is a Banach space. 
Then, we introduce the system
\eqn
		\Gamma(t) & = & e^{it\opDelta_\pm}\Gamma_0 \, - \, i \mu 
        \int_0^t ds \, e^{i(t-s)\opDelta_\pm} \Xi(s)
        \label{int-pGP-sys-Wspace-1} \\
        \Xi(t) & = & \opB \, e^{it\opDelta_\pm}\Gamma_0 \, - \, i \mu 
        \int_0^t ds \, \opB \, e^{i(t-s)\opDelta_\pm} \Xi(s) \,, 
        \label{int-pGP-sys-Wspace-2} 
\eeqn
which is formally equivalent to the system (\ref{int-pGP-sys-1}), (\ref{int-pGP-sys-2}).

Our main result in this section is the following existence and uniqueness theorem,
which corresponds to part (i) in Theorem \ref{thm-main-0}.

\begin{theorem}\label{thm-locwp-main-1}
Assume that $\alpha\in\alphaset(d,p)$ where $d\geq1$ and $p\in\{2,4\}$. 
Then, the following holds. For every $\Gamma_0\in\cH_{{\xi_1}}^\alpha$, there
exist constants $T>0$ and $0<\xi_2\leq\xi_1$ 
such that for $t\in I=[0,T]$, there exists a unique  
solution $(\Gamma(t),\Xi(t))$
of the system (\ref{int-pGP-sys-Wspace-1}), (\ref{int-pGP-sys-Wspace-2}), in the space $\Wspace$.
Moreover, the relation
\eqn
	\Xi(t)=\opB\Gamma(t)\in L^1_{t \in I}\cH_{\xi_2}^\alpha \,,
\eeqn
holds for the solution, and the component $\Gamma(t)$ satisfies the
$p$-GP hierarchy (\ref{seclwp-pGP}), 
\eqn\label{eq-GPhier-thm-1}
	i\partial_t\Gamma \, + \, \opDelta_\pm\Gamma \, - \,
	\mu \, \opB \Gamma \, = \, 0\,,
\eeqn
with initial condition $\Gamma(0)=\Gamma_0\in\cH_{\xi_1}^\alpha$.

Moreover, there exists a constant $C(T,d,p,\xi_1,\xi_2)$ such that
\eqn
	\| \, \opB \Gamma \, \|_{L^1_{t\in I}\cH_{\xi_2}^\alpha} \, \leq \,
	C(T,d,p,\xi_1,\xi_2) \, \| \, \Gamma_0 \, \|_{\cH_{\xi_1}^\alpha}
\eeqn
holds.
\end{theorem}

We note that our local existence and uniqueness result differs
than the one proven in \cite{esy1,esy2} in that performing the contraction principle on the space
$\Wspace$ presumes finiteness of $\| \, \opB \Gamma \, \|_{L^1_{t\in I}\cH_{\xi_2}^\alpha}$.
However,
\footnote{We thank B. Schlein for calling our attention to this fact.}  
it is a priori conceivable that there exist solutions of the $p$-GP hierarchy 
$\Gamma(t)\in \cH_{\xi_2}^\alpha$ for which $\| \, \opB \Gamma \, \|_{L^1_{t\in I}\cH_{\xi_2}^\alpha}$
is not finite.
We note that finiteness of $\| \, \opB \Gamma \, \|_{L^1_{t\in I}\cH_{\xi_2}^\alpha}$ is an
essential element of this analysis, and has been assumed without proof in \cite{klma} as a key
ingredient.

It was previously unknown whether this condition is generally satisfied
for solutions of the GP hierarchy. 
Using Theorem \ref{thm-locwp-main-1}, we can prove that 
$\| \, \opB \Gamma \, \|_{L^1_{t\in I}\cH_{\xi_2}^\alpha}<\infty$ is a sufficient
and necessary condition for a solution $\Gamma(t)\in\cH_{\xi_2}^\alpha$ of (\ref{seclwp-pGP}),
with initial condition $\Gamma(0)\in\cH_{\xi_1}^\alpha$, to be unique.  
Accordingly, we arrive at the following result which corresponds to part (ii)
in Theorem \ref{thm-main-0}.

\begin{theorem}\label{thm-locwp-main-1-L1bd}
Assume that $\alpha\in\alphaset(d,p)$ where $d\geq1$ and $p\in\{2,4\}$.   
Assume moreover that for $\Gamma_0\in\cH_{{\xi_1}}^\alpha$ there exist constants $T>0$ 
and $0<\xi_2\leq\xi_1$ such that for $t\in I=[0,T]$, there exists a  
solution $\Gamma(t)$ of the $p$-GP hierarchy (\ref{seclwp-pGP})
in the space $L^\infty_{t\in I}\cH_{\xi_2}^\alpha$.

Then, the solution $\Gamma(t)\in L^\infty_{t\in I}\cH_{\xi_2}^\alpha$ 
is {\em unique} if and only if 
$\| \, \opB \Gamma \, \|_{L^1_{t\in I}\cH_{\xi}^\alpha}<\infty$ 
holds for some $\xi>0$.

If the latter holds, then in fact,
there exists a constant $C(T,d,p,\xi_1,\xi_2)$ such that
\eqn\label{eq-Strichartz-Wspace-1-1}
	\| \, \opB \Gamma \, \|_{L^1_{t\in I}\cH_{\xi_2}^\alpha} \, \leq \,
	C(T,d,p,\xi_1,\xi_2) \, \| \, \Gamma_0 \, \|_{\cH_{\xi_1}^\alpha}
\eeqn
is satisfied. 
\end{theorem}

We first prove Theorem \ref{thm-locwp-main-1}.
\prf
We make the key observation that the fixed point equation
(\ref{int-pGP-sys-Wspace-2}) determining the component $\Xi(t)$ of the 
desired solution $(\Gamma(t),\Xi(t))$ is self-contained, and independent
of the component $\Gamma(t)$.
Therefore, we can invoke the Picard fixed point principle on the 
space $L^1_{t\in I}\cH_{\xi_2}^\alpha$ (the second factor of the product space
$\Wspace$) to first 
find a unique $\Xi\in L^1_{t\in I}\cH_{\xi_2}^\alpha$ solving (\ref{int-pGP-sys-Wspace-2}).
 
\noindent\underline{\em (1) Existence and uniqueness of the component $\Xi(t)$}

We recall that the fixed point equation for $\Xi(t)$ is given by
\eqn\label{eq-opBGamma-fixedpt-1}
	\Xi(t) \, = \, \opB e^{it\opDelta_\pm} \Gamma_0 \, - \, i 
	\mu \, \int_0^t \, ds \, \opB e^{i(t-s)\opDelta_\pm}\Xi(s)\,,
\eeqn
in the space $L^1\cH_{\xi_2}^\alpha$.

We define, for an admissible sequence of density matrices 
$\widetilde\Gamma=(\widetilde\gamma^{(k)})_{k\in\N}$, 
\eqn\label{eq-Duh-j-def-1}
	\lefteqn{
	\duh_j(\widetilde\Gamma)^{(k+1)}(t) 
	}
	\nonumber\\
	& := & (-i\mu)^j\int_0^t dt_1 \cdots \int_0^{t_{j-1}}dt_j
	e^{i(t-t_1)\Delta_\pm^{(k+1)}}B_{k+2}e^{i(t_1-t_2)\Delta_\pm^{(k+2)}}
	\nonumber\\
	&&\quad\quad\quad\quad\quad\quad
	B_{k+3} \cdots
	\cdots B_{k+j+1} e^{i t_j \Delta_\pm^{(k+j+1)}} \widetilde\gamma^{(k+j+1)}(t_j) \,.
\eeqn
Then, any solution of (\ref{eq-opBGamma-fixedpt-1}) satisfies the fixed point equation
(obtained from iterating the Duhamel formula $k$ times for the $k$-th component of $\opB\Gamma$)
\eqn
	(\opB\Gamma)^{(k)}(t) \, = \, \sum_{j=1}^{k-1}B_{k+1}\duh_j(\Gamma_0)^{(k+1)}(t)
	\, + \, B_{k+1}\duh_{k}(\opB\Gamma)^{(k+1)}(t) \,.
\eeqn
To formulate a Picard-type fixed point argument, we define
\eqn
	\Phi(\Xi)=( \Phi(\Xi)^{(k)})_{k\in\N}
\eeqn
where the $k$-th component is given by
\eqn\label{eq-PhiGamma-fixed-pt}
	\Phi(\Gamma)^{(k)}(t) \, = \, \sum_{j=1}^{k-1}B_{k+1}\duh_j(\Gamma_0)^{(k+1)}(t)
	\, + \, B_{k+1}\duh_{k}(\Xi)^{(k+1)}(t) \,.
\eeqn
Similarly as in \cite{chpa}, we use different approaches when $d\geq2$ and when $d=1$.
In dimension $d=1$, and for both the cubic and quintic GP hierarchy, 
we use a spatial a priori bound as in \cite{chpa}
where we refer for details.  

In dimensions $d\geq2$, we apply the 
Klainerman-Machedon spacetime bounds similarly to 
\cite{klma} and \cite{kiscst}.  
This is explained in detail in 
the Appendix.
 
We first consider the case $d\geq2$. 

In the sequel, we will often abbreviate
\eqn
	H^\alpha_k \, := \, H^\alpha(\R^{dk}\times \R^{dk}) \,.
\eeqn 
We invoke Propositions \ref{prp-spacetime-bd-1} and 
\ref{prp-spacetime-bd-2} in the Appendix. 
They generalize the  spacetime bounds and ``board game"
arguments developed in \cite{klma}.
Proposition \ref{prp-spacetime-bd-2}  implies that for $\Gamma_0=(\gamma_0^{(k)})_{k\in\N}$,
\eqn
	\lefteqn{
	\| \, \sum_{j=1}^{k-1}B_{k+1}\duh_j(\Gamma_0)^{(k+1)}(t) \, \|_{L^1_{t\in I}H^\alpha_k}
	}
	\nonumber\\
	& < &  k C^k \sum_{j=1}^{k-1} (c T)^{\frac {(j+1)}2} \|\gamma^{(k+j+1)}_0\|_{H^\alpha_{k+j+1}}
	\\
	& < & k  (C \xi_1^{-1})^{k} \sum_{j=1}^{k-1} (c T \xi_1^{-2})^{\frac {(j+1)}2} 
	\xi_1^{k+j+1} \|\gamma^{(k+j+1)}_0\|_{H^\alpha_{k+j+1}}
	\\
	& < & (c T \xi_1^{-2})  k  (C \xi_1^{-1})^{k} \sum_{j=1}^{k-1}  \xi_1^{k+j+1} 
	\|\gamma^{(k+j+1)}_0\|_{H^\alpha_{k+j+1}} \,.
	\label{eq-Duhexp-aux-1}
\eeqn
Therefore,
\eqn
	\lefteqn{
	\sum_{k=1}^\infty \xi_2^{k}\| \, \sum_{j=1}^{k-1}B_{k+1}\duh_j(\Gamma_0)^{(k+1)}(t) \, 
	\|_{L^1_{t\in I}H^\alpha_{k}}
	}
	\nonumber\\  
	& < & (c T \xi_1^{-2}) 
	\sum_{k=1}^\infty k  \Big(C \frac{\xi_2}{\xi_1}\Big)^k \sum_{j=1}^{k-1}  \xi_1^{k+j+1} 
	\|\gamma^{(k+j+1)}_0\|_{H^\alpha_{k+j+1}}
	\\
	& < & (c T \xi_1^{-2}) 
	\sum_{k=1}^\infty k \Big(C \frac{\xi_2}{\xi_1}\Big)^k
	\sum_{\ell=1}^{2k}  \xi_1^{\ell} \|\gamma^{(\ell)}_0\|_{H^\alpha_\ell}
	\\
	& < & (c T \xi_1^{-2})  
	\sum_{k=1}^\infty k \big(C \eta \big)^k
	\|\Gamma_0\|_{\cH_{\xi_1}^\alpha}
	\\
	& < & (c T \xi_1^{-2}) 
	\|\Gamma_0\|_{\cH_{\xi_1}^\alpha}
	\label{eq-Duhexp-aux-2}
\eeqn
for $\xi_2=\eta\xi_1$, with $0<\eta\leq1$ sufficiently small.

This implies that, for $I=[0,T]$, and any $T>0$,
\eqn
	\Big( \, \sum_{j=1}^{k-1}B_{k}\duh_j(\Gamma_0)^{(k+1)}(t) \, \Big)_{k\in\N}
	\, \in \, L^1_{t\in I}\cH^\alpha_{\xi_2} \,,
\eeqn
provided that $\Gamma_0\in\cH_{\xi_1}^\alpha$, and $\xi_2=\eta\xi_1$ with $\eta>0$ sufficiently small.

Our next step is to prove that $\Phi$ is a contraction on $L^1_{t\in I}\cH^\alpha_{\xi_2}$. 
To this end, we use the bound
\eqn
	\|\Phi(\Xi_1)^{(k)} -\Phi(\Xi_2)^{(k)} \|_{L_{t\in I}^1H^\alpha_k}
	\, \leq \, k \, (C T)^{\frac{k}{2}} \| \Xi_1^{(2k)} - \Xi_2^{(2k)} \|_{L^1_{t\in I}H^\alpha_{2k}}
\eeqn
which is proven in Proposition \ref{prp-spacetime-bd-2} in the Appendix. 

We obtain
\eqn\label{eq-PhiGamma-contr-1}
	\lefteqn{
	\|\Phi(\Xi_1)-\Phi(\Xi_2)\|_{L^1_{t\in I}\cH_{\xi_2}^\alpha}
	}
	\nonumber\\
	& =  &
	\sum_{k=1}^\infty\xi_2^k
	\|\Phi(\Xi_1)^{(k)} -\Phi(\Xi_2)^{(k)} \|_{L_{t\in I}^1H^\alpha_k}
	\nonumber\\
	&\leq& 
	\sum_{k=1}^\infty k \, (C T \xi_2^{-2})^{\frac{k}{2}} 
	\xi_2^{2k} \, \| \Xi_1^{(2k)} - \Xi_2^{(2k)} \|_{L^1_{t\in I}H^\alpha_{2k}}
	\nonumber\\
	&\leq& 
	\sup_{k\geq1}\{ \, k \, (C T \xi_2^{-2})^{\frac{k}{2}} \, \} \, \sum_{k=1}^\infty 
	\xi_2^{2k} \, \| \Xi_1^{(2k)} - \Xi_2^{(2k)} \|_{L^1_{t\in I}H^\alpha_{2k}}
	\nonumber\\
	&\leq&
	( C T \xi_2^{-2})^{\frac12} \,  
	\| \Xi_1 - \Xi_2 \|_{L^1_{t\in I}\cH_{\xi_2}^\alpha}\,,
\eeqn
for $T>0$ sufficiently small.
This implies that $\Phi$ is a contraction on $L^1_{t\in I}\cH^\alpha_{\xi_2}$ 
provided that $T$ is sufficiently small.

Consequently, for a every initial condition $\Gamma_0\in\cH_{\xi_1}^\alpha$,
there exist $T>0$ and $\xi_2=\eta\xi_1$ with $0<\eta\leq1$
such that there exists a unique solution $\Xi(t)$ of  (\ref{eq-opBGamma-fixedpt-1})  
in $L^1_{t \in I}\cH^\alpha_{\xi_2}$, 

Repeating the above arguments, we find that this solution satisfies 
\eqn\label{eq-Strichartz-cH-1}
	\| \, \Xi  \, \|_{L^1_{t\in I}\cH_{\xi_2}^\alpha}
	&\leq&(c T \xi_1^{-2}) 
	\|\Gamma_0\|_{\cH_{\xi_1}^\alpha}
	\, + \, ( C T \xi_2^{-2} )^{\frac12} \,  
	\| \, \Xi \, \|_{L^1_{t\in I}\cH_{\xi_2}^\alpha} \,.
\eeqn
Hence, in particular,
\eqn\label{eqn-BGamma-spacetime-bd-1}
	\| \, \Xi \, \|_{L^1_{t\in I}\cH_{\xi_2}}
	&\leq&\frac{c T \xi_1^{-2} }{1-(c T \xi_2^{-2} )^{\frac 12}}
	\|\Gamma_0\|_{\cH_{\xi_1}^\alpha} \, 
\eeqn
for sufficiently small $T>0$.

\noindent\underline{\em (2) Determination and uniqueness of the component $\Gamma(t)$}

Given the solution $\Xi$ in $L^1_{t \in I}\cH^\alpha_{\xi_2}$ obtained from
the contraction argument (\ref{eq-PhiGamma-contr-1}), the right hand
side of the integral equation (\ref{int-pGP-sys-Wspace-1}) is completely determined
by $\Gamma_0$ and $\Xi(s)$ where $s\in I$.

Accordingly,  $\Gamma(t)$ is given by
\eqn\label{int-pGP-sys-Wspace-1-2}
	\Gamma(t) \, = \, e^{it\opDelta_\pm}\Gamma_0 \, - \, i \mu \int_0^t ds \, e^{i(t-s)\opDelta_\pm}
	\Xi(s) \, 
\eeqn
where the solution $\Xi(s)$ found in step {\em (1)} above has been substituted into the integral. Evidently,
\eqn\label{eq-Gamma-Duh-opBGamma-1}
	\| \Gamma(t) \|_{ \cH_{\xi_2}^\alpha} 
	& \leq & \| \Gamma_0 \|_{\cH_{\xi_2}^\alpha}
	\, + \, \|\int_0^t  ds \,   e^{i(t-s)\opDelta_\pm}\Xi (s)\|_{\cH_{\xi_2}^\alpha}
	\nonumber\\
	& \leq & \| \Gamma_0 \|_{\cH_{\xi_2}^\alpha}
	\, + \, \int_0^t  ds \, \| \Xi (s)\|_{\cH_{\xi_2}^\alpha}
	\nonumber\\
	& \leq & \| \Gamma_0 \|_{\cH_{\xi_2}^\alpha}
	\, + \, \| \Xi(s) \|_{L^1_{s\in I}\cH_{\xi_2}^\alpha} \,,
\eeqn
using the unitarity of $e^{i(t-s)\opDelta_\pm}$ with respect to $\cH_{\xi_2}^\alpha$.
By (\ref{eqn-BGamma-spacetime-bd-1}), we conclude that the last line is bounded,
hence $\Gamma(t)$ determined by (\ref{int-pGP-sys-Wspace-1-2}) 
indeed lies in $L^\infty_{t\in I}\cH_{\xi_2}^\alpha$.

To prove uniqueness, we assume that $(\Gamma_1(t),\Xi_1(t))$ and $(\Gamma_2(t),\Xi_2(t))\in\Wspace$ 
correspond to solutions of the system (\ref{int-pGP-sys-Wspace-1}), (\ref{int-pGP-sys-Wspace-2}) to
the same initial condition $\Gamma_1(0)=\Gamma_2(0)\in\cH_{\xi_1}^\alpha$, for $t\in[0,T]$. 
Then, linearity implies that 
$(\Gamma_1(t)-\Gamma_2(t),\Xi_1(t))-\Xi_2(t))\in\Wspace$ 
solves the system (\ref{int-pGP-sys-Wspace-1}), (\ref{int-pGP-sys-Wspace-2}) 
with initial condition $\Gamma_1(0)-\Gamma_2(0)=(0)_{k\in\N}$.
Hence, (\ref{eq-Gamma-Duh-opBGamma-1}) implies that
\eqn
	\| \Gamma_1(t)-\Gamma_2(t) \|_{ \cH_{\xi_2}^\alpha}
	& \leq & \Big( \, 1 \,+ \, \frac{c T \xi_1^{-2}}{1-(c T \xi_2^{-2})^{\frac 12}} \, \Big) \,
	\|\Gamma_1(0)-\Gamma_2(0)\|_{\cH_{\xi_1}^\alpha} 
	\nonumber\\
	&=&0\,.
\eeqn
This implies that $\Gamma_1(t)=\Gamma_2(t)$, for all $t<T$. 
This proves local existence and uniqueness of solutions for the 
system (\ref{int-pGP-sys-Wspace-1}), (\ref{int-pGP-sys-Wspace-2}).

Next, we observe that applying $\opB$ to $\Gamma(t)$ produces the rhs of
the fixed point equation (\ref{eq-opBGamma-fixedpt-1}) for $\Xi(t)$. Therefore,
we immediately find that 
\eqn\label{eq-XiopBGammadiff-1}
	\Xi \, = \,  \opB \Gamma  \; \; \; \in \, L^1_{t\in I}\cH_{\xi_2}^\alpha \,.
\eeqn
Finally, it is easy to verify for the unique solution 
$(\Gamma(t),\Xi(t))\in\Wspace$ of  (\ref{int-pGP-sys-Wspace-1}), (\ref{int-pGP-sys-Wspace-2}) that
\eqn
	i\partial_t \Gamma(t) \, + \, \opDelta_\pm \Gamma(t) \, = \, \mu \, \Xi(t) \,.
\eeqn
Therefore, using (\ref{eq-XiopBGammadiff-1}),
\eqn
	\| \, i\partial_t \Gamma(t) \, + \, \opDelta_\pm \Gamma(t) 
	\, - \, \mu \, \opB \Gamma(t) \, \|_{L^1_{t\in I}\cH_{\xi_2}^\alpha} \, = \, 0 \,,
\eeqn
which implies (\ref{eq-GPhier-thm-1}).

For the quintic GP hierarchy, all steps of the above proof can be adopted with minor modifications. 
A key difference is the fact that (\ref{eq-Gamma-Duh-opBGamma-1}) is replaced by 
\eqn\label{eq-Gamma-Duh-opBGamma-quintic-1}
	\| \Gamma(t) \|_{ \cH_{\xi_2}^\alpha} & \leq & \| \Gamma_0 \|_{\cH_{\xi_2}^\alpha}
	\, + \,  \| \opB\Gamma \|_{L^1_{t\in I}\cH_{\xi_2}^\alpha}
	\nonumber\\
	 & \leq & \Big( \, 1 \,+ \, \frac{c T \xi_1^{-4}}{1-(c T \xi_2^{-4})^{\frac 12}} \, \Big) \,
	\|\Gamma_0\|_{\cH_{\xi_1}^\alpha} \,.
\eeqn
This concluded the proof of Theorem \ref{thm-locwp-main-1} for $d\geq2$.


%
In the case $d=1$, we can straightforwardly adapt the proof given in \cite{chpa}
of the uniqueness of solutions of the quintic GP hierarchy
in $d=1$. The spacetime bounds of Proposition \ref{prp-spacetime-bd-1}
is not available in $d=1$ since it would produce divergent bounds.
However, the spatial bounds in $d=1$ proven in \cite{chpa} apply for both the
cubic and the quintic GP hierarchy, under the assumption that $\alpha>\frac12$.

The result is that we get a factor $t$ instead of $t^{\frac12}$, in all of the bounds
found above for the cubic GP hierarchy that 
produced a factor $t^{\frac12}$. Accordingly, we find 
\eqn\label{eq-Gamma-Duh-opBGamma-1D-1}
	\| \Gamma(t) \|_{ \cH_{\xi_2}^\alpha} & \leq & 
	\Big( \, 1 \,+ \, \frac{c T^2 \xi_1^{-2}}{1-c T \xi_2^{-1} } \, \Big) \,
	\|\Gamma_0\|_{\cH_{\xi_1}^\alpha}  
\eeqn
for the cubic GP hierarchy instead of (\ref{eq-Gamma-Duh-opBGamma-1}), and 
\eqn\label{eq-Gamma-Duh-opBGamma-quintic-1D-1}
	\| \Gamma(t) \|_{ \cH_{\xi_2}^\alpha}  
	 & \leq & \Big( \, 1 \,+ \, \frac{c T^2 \xi_1^{-4}}{1- c T  \xi_2^{-2} } \, \Big) \,
	\|\Gamma_0\|_{\cH_{\xi_1}^\alpha}  
\eeqn
for the quintic GP hierarchy instead of (\ref{eq-Gamma-Duh-opBGamma-quintic-1}), respectively.
Hence, we obtain local wellposedness for sufficiently small $T>0$.

In conclusion, we have proved the existence and uniqueness of solutions 
of the GP hierarchy (\ref{seclwp-pGP}) in
the space $\Wspace$ defined in (\ref{def-Wspace-1}).
\endprf

It is now straightforward to prove Theorem \ref{thm-locwp-main-1-L1bd}.

\prf
$"\Rightarrow"$. We first prove that uniqueness of $\Gamma\in L^\infty_{t\in I}\cH_{\xi_2}^\alpha$
implies  $\| \opB\Gamma \|_{L^1_{t \in I} \cH^\alpha_{\xi}}<\infty$ for
some $\xi>0$.
Let $t'\in I_1=[0,T_1]$.
Assume $\Gamma(t+t')$ is the unique solution in $\cH_{\xi_2}^1$ 
of the $p$-GP hierarchy for the initial condition $\Gamma(t)\in \cH_{\xi_1}^\alpha$,
for $t'\in I_1$, and $0<\xi_1<1$ chosen appropriately. 

We may also use $\Gamma(t)$ as an initial condition for the system 
(\ref{int-pGP-sys-Wspace-1}), (\ref{int-pGP-sys-Wspace-2}) in $\mathfrak{W}_{\xi_2}^\alpha(I_2)$.
Thereby, we obtain a unique solution $( \Gamma'(t+t') , \Xi(t+t') $ in $\mathfrak{W}_{\xi_2}^\alpha(I_2)$, for
$t' \in I_2 = [0,T_2]$, for $T_2>0$ and $0<\xi_2\leq\xi_1$ chosen small enough.

Let $I:=[0,\min\{T_1,T_2\}]$.
From Theorem  \ref{thm-locwp-main-1}, we infer that $\Gamma'(t+t')$ is also an
element of $\cH_{\xi_2}^\alpha$, for $t\in I$, and solves the GP
hierarchy equation with initial condition $\Gamma'(t)=\Gamma(t)$ for $t'=0$.

Since by assumption,  $\Gamma(t+t')$ is the unique solution of the GP hierarchy
in $\cH_{\xi_2}^\alpha$, solving the same GP hierarchy equation, and with the same initial condition
$\Gamma(t)$ at $t'=0$.
Thus, $\Gamma(t+t')=\Gamma'(t+t')$.

But since the solution $( \Gamma'(t+t') , \Xi(t+t') )$ in $\Wspace$  for the system 
(\ref{int-pGP-sys-Wspace-1}), (\ref{int-pGP-sys-Wspace-2}) has the property that 
$\| \opB\Gamma'(t + . ) \|_{L^1_{t'\in I} \cH^\alpha_{\xi_2}}
\leq C(T,d,p,\xi_1,\xi_2) \, \| \Gamma(t) \|_{\cH_{\xi_1}^\alpha}$
is bounded, we conclude that also, 
$\| \opB\Gamma(t + . ) \|_{L^1_{t'\in I} \cH^\alpha_{\xi_2}}
\leq C(T,d,p,\xi_1,\xi_2) \, \| \Gamma(t) \|_{\cH_{\xi_1}^\alpha}$.

$"\Leftarrow"$. Next, we prove that $\| \opB\Gamma \|_{L^1_{t \in I} \cH^\alpha_{\xi}}<\infty$ for
some $\xi>0$ implies uniqueness of 
$\Gamma\in L^\infty_{t\in I}\cH_{\xi_2}^\alpha$, for some $\xi_2$.
We may set $\xi=\xi_2$ by choosing $\min\{\xi,\xi_2\}$.

We note that since
$\Gamma\in L^\infty_{t\in I}\cH_{\xi_2}^\alpha$ and 
$\| \opB\Gamma \|_{L^1_{t \in I} \cH^\alpha_{\xi}}<\infty$, it is clear that
$(\Gamma,\opB\Gamma)\in\Wspace$. Accordingly, Theorem \ref{thm-locwp-main-1}
implies that $\Gamma$ is the unique solution in $\Wspace$ for the
initial condition $\Gamma(0)=\Gamma_0\in\cH_{\xi_1}^\alpha$ for which 
$\| \opB\Gamma \|_{L^1_{t \in I} \cH^\alpha_{\xi}}<\infty$ is satisfied.

Finally, Theorem \ref{thm-locwp-main-1} implies that whenever 
$\| \opB\Gamma \|_{L^1_{t \in I} \cH^\alpha_{\xi}}<\infty$ holds for some $\xi>0$, 
the stronger estimate (\ref{eq-Strichartz-Wspace-1-1}) is satisfied.
\endprf
  
$\;$\\ \\
 
\section{On the global existence and uniqueness of solutions for defocusing GP hierarchies}
\label{sec-gwp-GP-1}

It is proved for the $d=2,3$ defocusing cubic case in \cite{esy1,esy2,kiscst}, 
and for the $d=1,2$ defocusing quintic 
case in \cite{chpa}, that whenever $\Gamma(t)=(\gamma^{(k)}(t))_{k\in\N}$ is  obtained from the
$N\rightarrow\infty$ limit of a bosonic $N$-particle BBGKY hierarchy with 
repulsive interactions, 
\eqn\label{eq-GPsol-defoc-apriorien-1}
	\tr \big( \, S^{(k,1)} \, \gamma^{(k)}_0 \, \big) \, < \, C^k
\eeqn
for some constant $C$, then
\eqn\label{eq-GPsol-defoc-apriorien-2}
	\tr \big( \, S^{(k,1)} \, \gamma^{(k)}(t) \, \big) \, < \, C_0^k
\eeqn
with $C_0$ independent of $t\geq0$.

This follows from energy conservation in the $N$-particle system
of which the GP hierarchy is the $N\rightarrow\infty$ limit.

We first claim that this implies that $\Gamma(t)\in\cH_{\xi_1}^1$ for some $\xi_1>0$,
and all $t\in\R$.

To see this, we consider a fixed $k$. Let $\gamma^{(k)}$ be non-negative,
normalized trace class, 
$\tr(\gamma^{(k)})=1$, and hermitean.
Then, we have that
\eqn
	\gamma^{(k)}(\ux_k;\ux_k') \, = \, \sum_j \lambda_j \, 
	|\psi_j(\ux_k')\rangle\langle\psi_j(\ux_k)|
\eeqn
for an orthogonal basis $\psi_j$ of $L^2(\R^{dk})$ with $\lambda_j\geq0$ and $\sum\lambda_j=1$.
Then,
\eqn
	\| \, \gamma^{(k)} \, \|_{H^1} 
	& = & \sum_{j,j'} \lambda_j \, \lambda_{j'}
	\big|\bra \, \langle\nabla_{\ux_k}\rangle\psi_j \, 
	\big| \,  \langle\nabla_{\ux_k}\rangle\psi_{j'} \, \ket \, \big|^2
	\\
	&\leq&
	\Big(\sum_j \lambda_j \,  
	\big\| \, \langle\nabla_{\ux_k}\rangle\psi_j \, \big\|^2 \, \Big)^2
	\\
	&=&
	\Big( \, \tr \big( \, S^{(k,1)} \, \gamma^{(k)} \, \big) \, \Big)^2 \,.
\eeqn
Thus, for a solution $\gamma^{(k)}(t)$ of the (cubic or quintic) GP hierarchy
with initial condition satisfying (\ref{eq-GPsol-defoc-apriorien-1}), we have that
\eqn 
	\| \, \gamma^{(k)}(t) \, \|_{H^1}
	\, < \, C_0^k
\eeqn
with $C_0$ independent of $t$.
Thus, for $\xi_1$ sufficiently small, we obtain the a priori bound
\eqn 
	\|\Gamma(t)\|_{\cH_{\xi_1}^1}
	\, \leq \, \Big( \, \sum_{k=1}^\infty (C_0\xi)^{ 2k} \, \Big)^{\frac12} \, < \, \infty \, ,
\eeqn
for all $t\in\R$.

This implies the existence of a solution $\Gamma(t)\in\cH_{\xi_1}^1$ globally in time.
The key question is whether $\Gamma(t)\in \cH_{\xi_1}^1$ is the unique solution 
of the $p$-GP hierarchy with initial condition $\Gamma_0\in\cH_{\xi_1}^1$.
To this end, we prove the following uniqueness result.

\begin{theorem}\label{thm-defocGP-globalwp-1}
Assume that $1\leq d\leq3$ for $p=2$, and 
in $1\leq d\leq 2$ for $p=4$ such that $1 \in\alphaset(d,p)$.
Assume that $\Gamma(t)\in\cH_{\xi_1}^1$,  $t\in\R$,
is a solution of the defocusing $p$-GP hierarchy for the initial condition
$\Gamma(0)=\Gamma_0$, and 
satisfies the a priori bound $\|\Gamma(t)\|_{\cH_{\xi_1}^1}<c_0<\infty$
for $t\in\R$. 
Moreover, assume that  for every
sufficiently short open interval $I\in\R$, 
there exists a constant $\xi=\xi(I)$ with $0<\xi\leq \xi_1$ such that 
$\|\opB\Gamma\|_{L^1_{t\in I}\cH_{\xi}^1}<\infty$.

Then, there exists a constant $\xi_2=\xi_2(\xi_1)$ depending only on $\xi_1$, 
and a constant $T_0=T_0(d,p,\xi_1,\xi_2)$ such that 
the pair $(\Gamma,\opB\Gamma)$ is the unique solution
of the system  (\ref{int-pGP-sys-Wspace-1}), (\ref{int-pGP-sys-Wspace-2})
associated to the defocusing ($\mu=+1$) $p$-GP hierarchy with initial condition
$\Gamma(0)=\Gamma_0\in\cH_{\xi_1}^1$, in the space
\eqn
	\Wspaceglob \, := \, \bigcup_{I\subset \R,|I|<T_0}\W1space \,,
\eeqn 
for $0<\xi_2\leq\xi_1$ sufficiently small.
Hence in particular,
\eqn
	\|\opB\Gamma\|_{L^1_{t\in I}\cH_{\xi_2}^1} \, \leq \, C(T_0,d,p,\xi_1,\xi_2) \, c_0
\eeqn
for all $I\in\R$ with $|I|<T_0$.
\end{theorem}

\prf   
Assume that $I'\subset\R$ is such that $\|\opB\Gamma\|_{L^1_{t\in I'}\cH_{\xi}^1}<\infty$.
Let $I:=[\tau,T]\subset I'$
Then,  Theorem \ref{thm-locwp-main-1-L1bd} implies that in fact, there exists a constant $\xi_2\leq\xi_1$
such that
\eqn
	\|\opB\Gamma\|_{L^1_{t\in I }\cH_{\xi_2}^1} \, < \, C_0 \, \| \, \Gamma(\tau) \, \|_{\cH_{\xi_1}^1}
	\, < \, C_0 \, c_0 \,,
\eeqn
for a constant $C_0=C_0(|I|,d,p,\xi_1,\xi_2)$.

This estimate holds independently of the value of  $\tau\in\R$. Therefore, we can cover $\R$
with intervals $\R=\cup_j I_j$ where $I_j=[\tau_j,\tau_{j+1}]$, $j\in\Z$, and $|I_j|=|I|=: T_0$.

Assume that the initial time $t=0\in I_0$. Then,  Theorem \ref{thm-locwp-main-1-L1bd} implies that
there exists a unique (forward and backward in time) solution  $(\Gamma,\opB\Gamma)$  
of the system  (\ref{int-pGP-sys-Wspace-1}), (\ref{int-pGP-sys-Wspace-2})
in $\mathfrak{W}_{\xi_2}^1(I_0)$ for the initial condition 
initial condition $\Gamma_0$.

By iteration, we use $\Gamma(\tau_{1})$ as the initial condition for $t\in I_1$,
and $\Gamma(\tau_0)$ as the terminal condition for $t\in I_{-1}$, and continue recursively,
thereby covering $\R=\cup I_j$.

In conclusion, we obtain that there exists a unique global in time solution
in  $\Wspaceglob$ with initial condition $\Gamma_0\in\cH_{\xi_1}^1$, 
under the assumptions of the theorem.
\endprf

\begin{remark}
As noted in Remark \ref{rem-BBGKYlim-1},  
every solution $\Gamma(t)\in\cH_{\xi_1}^1$ extracted by a diagonal argument 
from the $N\rightarrow\infty$ limit of the $N$-particle BBGKY hierarchies
(with repulsive interactions)
studied by Erd\"os-Schlein-Yau in \cite{esy1,esy2}, Kirkpatrick-Schlein-Staffilani in
\cite{kiscst}, and Chen-Pavlovi\'c in \cite{chpa}, satisfies 
$\| \, \opB \Gamma \, \|_{L^1_{t\in I}\cH_{\xi}^\alpha}<\infty$ 
for some $\xi>0$, and all sufficiently short intervals $I\subset\R$.
This follows a posteriori from the uniqueness of those solutions,
which has been established in these
works with independent methods, combined with Theorem \ref{thm-locwp-main-1-L1bd} in the work at hand.
Accordingly, $(\Gamma,\opB\Gamma)$
is automatically an element of $\Wspaceglob$ whenever  $\Gamma(t)$ is obtained from an $N\rightarrow\infty$
limit of the above noted type.  
\end{remark}

$\;$ \\ \\

\section{Lower bound on the blowup rates}
\label{sec-Proof-blowuprate-1}

In this section, we establish Theorem \ref{thm-blowuprate-L2crit-1}.
We adapt a standard proof given for $L^2$-critical focusing NLS to 
the GP hierarchy; see for instance \cite{ra}.
Let $p \in \{2,4\}$. 
Similarly as in (\ref{eq-pseudoconf-gamma-1}), one finds that 
the $p$-GP hierarchy is invariant under the rescaling 
\eqn\label{eq-pseudoconf-gamma-2}
	\lefteqn{
	\cS_{\lambda,t}\gamma^{(k)} (\tau,\ux_k;\ux_k')
	}
	\nonumber\\
	&&
	\, := \, 
	\frac{1}{(\lambda(t))^{4k/p}} \,  
	\gamma^{(k)}\Big( \, t +(\lambda(t))^{-2}\tau \, , \, (\lambda(t))^{-1}\ux_k \, ; \, 
	(\lambda(t))^{-1}\ux_k' \, \Big) \, 
\eeqn
If $\Gamma(t)=(\gamma^{(k)}(t))_{k\in\N}$
solves the $p$-GP hierarchy, then
$\cS_{\lambda,t}\Gamma=(\cS_{\lambda,t}\gamma^{(k)} )_{k\in\N}$
is also a solution of the $p$-GP hierarchy.
The proof can be straightforwardly adapted from the one given in Section \ref{sec-confinv-1}.

\noindent\underline{\em Proof of statement (a).}
\\
\\
Blowup in finite time means that there exists $T^*<\infty$ such that 
$\Av_{H^\alpha}(\Gamma(t))\rightarrow\infty$
as $t\rightarrow T^*$.
To prove a lower bound on the blowup rate, we may assume that 
$1<\Av_{H^\alpha}(\Gamma(t))<\infty$ at a fixed time $t$, and choose
\eqn
	\lambda(t) \, = \, (\Av_{H^\alpha}(\Gamma(t)))^{\frac{1}{2\alpha-d+\frac4p}} \, > \, 1 \,.
\eeqn
We note that $\frac{4}{d}<p<\frac{4}{d-2\alpha}$  implies that $2\alpha-d+\frac4p>0$.
Let 
\eqn
	S^{(k,\alpha)}_{\lambda(t)}
	\, := \, \prod_{j=1}^k\langle \, (\lambda(t))^{-1} \, \nabla_{x_j}\rangle^\alpha \,
	\langle \, (\lambda(t))^{-1} \, \nabla_{x_j'} \, \rangle^\alpha
\eeqn 
where $\langle b\nabla_{x}\rangle=\sqrt{1-b^2\Delta_{x}}$ for any $b\in\R$.
Clearly,
\eqn\label{eq-Sdot-quasiblowup-L2-aux-1}
	\lefteqn{
	\Big\| \, S^{(k,\alpha)} \cS_{\lambda,t}\gamma^{(k)} (\tau) \, \Big\|_{L^2_{\ux_k,\ux_k'}}
	}
	\nonumber\\
	& = &  (\lambda(t))^{k(d-\frac{4}{p})} \, \| \, (\, S^{(k,\alpha)}_{\lambda(t)} 
	\, \gamma^{(k)} \, ) (t+(\lambda(t))^{-2}\tau)\, \|_{L^2_{\ux_k,\ux_k'}} \,,
\eeqn
and
\eqn\label{eq-scaleSkalpha-bd-1}
	(\lambda(t))^{-2\alpha k} S^{(k,\alpha)} \, \leq \, 
	S^{(k,\alpha)}_{\lambda(t)}
	\, \leq \, S^{(k,\alpha)}
\eeqn
since we are assuming that $\lambda(t)>1$.

We define
\eqn
	\xi_<(\xi,t,\lambda) \, := \, \xi \, (\lambda(t))^{-\frac4p+d-2\alpha } 
	\, = \, \xi \, (\Av_{H^\alpha}(\Gamma(t)))^{-1}
\eeqn
and
\eqn\label{eq-xibigxi-id-1}
	\xi_>(\xi,t,\lambda) \, := \, \xi (\lambda(t))^{-\frac4p+d }
	\,  = \, \xi  (\Av_{H^\alpha}(\Gamma(t)))^{\frac{d-\frac4p}{2\alpha-d+\frac4p} }  \,.
\eeqn
Clearly, (\ref{eq-Sdot-quasiblowup-L2-aux-1}) and (\ref{eq-scaleSkalpha-bd-1}) imply that
\eqn\label{eq-sandwich-bd-1}
	\| \, \Gamma(t+(\lambda(t))^{-2}\tau) \, \|_{\cH_{\xi_<(\xi,t,\lambda)}^\alpha} 
	& \leq & 
	\|\, \cS_{\lambda,t}\Gamma (\tau) \, \|_{ \cH^\alpha_{\xi}}
	 \\
	&\leq&
	\| \, \Gamma(t+(\lambda(t))^{-2}\tau) \, \|_{\cH_{\xi_>(\xi,t,\lambda)}^\alpha} \,.
	\nonumber
\eeqn 
As a consequence of the definition of $\Av_{H^\alpha}(\Gamma(t))$,
it follows that for $\tau=0$,
\eqn
	0 \, < \, \| \, \Gamma(t  ) \, \|_{\cH_{\xi_<(\xi,t,\lambda)}^\alpha} \, < \, c
\eeqn
for any $0<\xi<1$.  

To ensure that $\|  \Gamma(t+(\lambda(t))^{-2}\tau)  \|_{\cH_{\xi_>(\xi,t,\lambda)}^\alpha}<c$, 
we use the fact that according to (\ref{eq-xibigxi-id-1}), $\xi_>(\xi,t,\lambda) $
can be made arbitrarily small by choosing $\xi$ small.

We note that our assumption $\frac{4}{d}<p<\frac{4}{d-2\alpha}$  implies that $2\alpha-d+\frac4p>0$ 
and $d-\frac4p>0$, so that the exponent on the rhs of (\ref{eq-xibigxi-id-1}) is positive. 
If blowup occurs, such that $\Av_{H^\alpha}(\Gamma(t))\rightarrow\infty$ as $t\nearrow T^*$,
the above considerations necessitate the choice of values of $\xi$ (whose reciprocal determines the energy scale)  
tending to zero as $t\nearrow T^*$.

Thus, for $\xi_1>0$ sufficiently small,
\eqn
	\|\, \cS_{\lambda,t}\Gamma (0) \, \|_{ \cH^\alpha_{\xi_1}}
	\, \leq \,
	\| \, \Gamma(t ) \, \|_{\cH_{\xi_>(\xi_1,t,\lambda)}^\alpha}
\eeqn
Due to Theorem  \ref{thm-locwp-main-1}, we may   pick $0<\xi_2=\eta\xi_1\ll\xi_1<1$, 
such that there exists a solution 
$\cS_{\lambda,t}\gamma^{(k)} (\tau)\in\cH_{\xi_2}^\alpha$ if
$\tau\in[0,\tau_{\max}]$, for $\tau_{\max}>0$ sufficiently small.

But this implies that
\eqn
	\| \, \Gamma(t+(\lambda(t))^{-2}\tau) \, \|_{\cH_{\xi_<(\xi_2,t,\lambda)}^\alpha}
	\, \leq \,
	\|\, \cS_{\lambda,t}\Gamma (\tau) \, \|_{ \cH^\alpha_{\xi_2}} \, < \, \infty
\eeqn
for $\tau\in[0,\tau_{\max}]$ so that there is no blowup if $\tau$ lies in that
interval. Therefore, the blowup time $T^*$ is bounded from below by
\eqn
	T^* \, > \, t+(\lambda(t))^{-2}\tau_{\max} \,,
\eeqn
and hence,
\eqn
	( \, \Av_{H^\alpha}(\Gamma(t)) \, )^{\frac12} \, = \, 
        \lambda(t)^{(\alpha-\frac{d}{2}+\frac{2}{p})} 
	\, > \, \frac{C}{|T^*-t|^{(2\alpha-d+\frac4p )/4}} \,.
\eeqn
This proves ($a$).

\noindent\underline{\em Proof of statement (b).}
\\
\\
It is easy to see that
\eqn\label{eq-Lp-rescale}
	\Big\| \, \cS_{\lambda,t}\gamma^{(k)} (\tau) \, \Big\|_{L^r_{\ux_k,\ux_k'}}
	\, = \, (\lambda(t))^{-2k(\frac2p-\frac{d}{r})} \, 
	\| \,  \gamma^{(k)} (t+(\lambda(t))^{-2}\tau)\, \|_{L^r_{\ux_k,\ux_k'}} \,,
\eeqn
which, in turn, implies that 
\eqn
	\|\, \cS_{\lambda,t}\gamma^{(k)} (0) \, \|_{\cL^r_\xi}
	&=&
	\sum_{k\geq1}\xi^k\Big\| \, \cS_{\lambda,t}\gamma^{(k)} (0) \, \Big\|_{L^r_{\ux_k,\ux_k'}} 
	\nonumber\\
	& = & \sum_{k\geq1} \xi^k \, (\lambda(t))^{-2k(\frac2p-\frac{d}{r})} \, 
	\| \,  \gamma^{(k)} (t)\, \|_{L^r_{\ux_k,\ux_k'}}
	\nonumber\\
	& = & \sum_{k\geq1} \left( \, \frac{\xi}{(\lambda(t))^{(\frac4p-\frac{2d}{r})}} \, \right)^{k} \, 
	\| \,  \gamma^{(k)} (t)\, \|_{L^r_{\ux_k,\ux_k'}} \,. \label{bigLp-lambda}
\eeqn
However \eqref{bigLp-lambda} is bounded for every $\xi<1$, if we choose
\eqn
	\lambda(t) \, = \, ( \, \Av_{L^r}(\Gamma(t)) \, )^{\frac{1}{\frac{4}{p}-\frac{2d}{r}}} \,.
\eeqn

Now we argue as in the proof of the part ($a$) by using 
the local well-posedness Theorem \ref{thm-locwp-main-1} 
to conclude that the $H^\alpha$ blowup time $T^*$ is bounded from below by
\eqn
	T^* \, > \, t+(\lambda(t))^{-2}\tau_{\max} \,.
\eeqn
Therefore 
\eqn
	( \, \Av_{L^r}(\Gamma(t)) \, )^{\frac12} \, = \, 
        \lambda(t)^{(\frac{2}{p}-\frac{d}{r})} 
	\, > \, \frac{C}{|T^*-t|^{\frac{1}{p} - \frac{d}{2r}}} \,.
\eeqn
Hence ($b$) is proved. 
\qed

\section{Proof of pseudoconformal invariance}
\label{sec-confinv-1}

In this section, we prove Theorem \ref{thm-pseudoconf-gamma-1}.
We recall the pseudoconformal transformations
\eqn\label{eq-pseudoconf-gamma-1}
	\lefteqn{
	{\mathcal P}\gamma^{(k)} (t,\ux_k;\ux_k')
	}
	\nonumber\\
	&&
	\, := \, 
	\frac{1}{(1+bt)^{dk}} \, e^{-i\frac{b(|\ux_k|^2-|\ux_k'|^2)}{1+bt}}
	\gamma^{(k)}\Big( \, \frac{1}{1+bt} \, , \, \frac{\ux_k}{1+bt} \, ; \, \frac{\ux_k'}{1+bt} \, \Big) \,,
\eeqn
for any $b\in\R\setminus\{0\}$.
Similarly as in the case of NLS, one can verify that
\eqn
	\lefteqn{
	(i\partial_t+ \Delta_{\ux_k}-\Delta_{\ux_k'}) {\mathcal P}\gamma^{(k)} (t,\ux_k;\ux_k')
	}
	\\
	&=&\frac{1}{(1+bt)^2}\frac{1}{(1+bt)^{dk}} \, e^{-i\frac{b(|\ux_k|^2-|\ux_k'|^2)}{1+bt}}
	\nonumber\\
	&&\quad\quad\quad
	((i\partial_t+ \Delta_{\ux_k}-\Delta_{\ux_k'})\gamma^{(k)})
	\Big( \, \frac{1}{1+bt} \, , \, \frac{\ux_k}{1+bt} \, ; \, \frac{\ux_k'}{1+bt} \, \Big) \,.
	\nonumber
\eeqn
Now we shall prove the pseudoconformal invariance of the quintic GP hieararchy when $d=1$.
In particular, we find that
\eqn
	\lefteqn{
	B_{j;k+1,k+2}^1{\mathcal P}\gamma^{(k+2)} (t,\ux_k;\ux_k')
	}
	\nonumber\\
	&=&
	\frac{1}{(1+bt)^{d(k+2)}} \, e^{-i\frac{b(|\ux_k|^2-|\ux_k'|^2)}{1+bt}}
	\int dx_{k+1} dx_{k+2} dx_{k+1}' dx_{k+2}'
	\nonumber\\
	&&\quad \delta(x_j-x_{k+1})\delta(x_j-x_{k+1}')
	\delta(x_j-x_{k+2})\delta(x_j-x_{k+2}')
	\nonumber\\
	&&\quad\quad
	\gamma^{(k)}\Big( \, \frac{1}{1+bt} \, , \, \frac{(\ux_k,x_{k+1},x_{k+2})}{1+bt} \, ; 
	\, \frac{(\ux_k',x_{k+1}',x_{k+2}')}{1+bt} \, \Big)
	\\
	&=&
	\frac{1}{(1+bt)^{2d}}\frac{1}{(1+bt)^{dk}} \, e^{-i\frac{b(|\ux_k|^2-|\ux_k'|^2)}{1+bt}} 
	\nonumber\\ 
	&&\quad\quad
	\gamma^{(k)}\Big( \, \frac{1}{1+bt} \, , \, \frac{(\ux_k,x_{j},x_{j})}{1+bt} \, ; 
	\, \frac{(\ux_k',x_{j},x_{j})}{1+bt} \, \Big) \,,
	\\
	&=&
	\frac{1}{(1+bt)^{2d}}\frac{1}{(1+bt)^{dk}} \, e^{-i\frac{b(|\ux_k|^2-|\ux_k'|^2)}{1+bt}} 
	\nonumber\\ 
	&&\quad\quad  
	B_{j;k+1,k+2}^1\gamma^{(k+2)} \Big( \, \frac{1}{1+bt} \, , \, \frac{\ux_k}{1+bt} \, ; 
	\, \frac{\ux_k'}{1+bt} \, \Big)
\eeqn
with $B_{j;k+1,k+2}=B_{j;k+1,k+2}^1-B_{j;k+1,k+2}^2$; in $B_{j;k+1,k+2}^2$, 
the variable $x_j$ in $B_{j;k+1,k+2}^1$ is replaced by $x_j'$. Notably, we have used that
\eqn
	e^{-i\frac{b((x_{k+1}^2+x_{k+2}^2)-(\ux_{k+1}'^2+x_{k+2}'^2))}{1+bt}}\Big|_{x_{k+1}=x_{k+2}=x_{k+1}'=x_{k+2}'=x_j}
	\, = \, 1 \,.
\eeqn
Thus, when $d=1$, we obtain
\eqn
	\lefteqn{
	\Big( \, (  i\partial_t+ \Delta_{\ux_k}-\Delta_{\ux_k'}) 
	{\mathcal P}\gamma^{(k)} - \mu \sum_{j=1}^k B_{j;k+1,k+2}
	{\mathcal P}\gamma^{(k+2)} \, \Big) (t,\ux_k;\ux_k')
	}
	\nonumber\\
	&=&\frac{1}{(1+bt)^2}\frac{1}{(1+bt)^{dk}} \, e^{-i\frac{b(|\ux_k|^2-|\ux_k'|^2)}{1+bt}}
	\nonumber\\
	&& 
	\Big(  (i\partial_t+ \Delta_{\ux_k}-\Delta_{\ux_k'})\gamma^{(k)}
	- \mu \sum_{j=1}^k B_{j;k+1,k+2}\gamma^{(k+2)} \, \Big)
	\Big( \, \frac{1}{1+bt} \, , \, \frac{\ux_k}{1+bt} \, ; \, \frac{\ux_k'}{1+bt}   \Big)  
	\nonumber\\
	&=&0 \,.
\eeqn 
This proves pseudoconformal invariance of the quintic GP hierarchy in dimension $d=1$.

For the cubic GP hierarchy, the operators
$B_{j;k+1,k+2}$ are replaced by operators $B_{j;k+1}$ which contract $x_j$, $x_j'$
only with $x_{k+1}$ and $x_{k+1}'$, 
\eqn
	(i\partial_t+ \Delta_{\ux_k}-\Delta_{\ux_k'})\gamma^{(k)}
	- \mu\sum_{j=1}^k B_{j;k+1}\gamma^{(k+1)} \, = \, 0 \,.
\eeqn 
The same considerations as above then produce
\eqn
	\lefteqn{
	\Big( \, (  i\partial_t+ \Delta_{\ux_k}-\Delta_{\ux_k'}) 
	{\mathcal P}\gamma^{(k)} -  \mu \sum_{j=1}^k B_{j;k+1}
	{\mathcal P}\gamma^{(k+1)} \, \Big) (t,\ux_k;\ux_k')
	}
	\nonumber\\
	&=&\frac{1}{(1+bt)^2}\frac{1}{(1+bt)^{dk}} \, e^{-i\frac{b(|\ux_k|^2-|\ux_k'|^2)}{1+bt}}
	\nonumber\\
	&& 
	\Big( \, (i\partial_t+ \Delta_{\ux_k}-\Delta_{\ux_k'})\gamma^{(k)}
	-  \mu \sum_{j=1}^k B_{j;k+1}\gamma^{(k+1)} \, \Big)
	\Big( \, \frac{1}{1+bt} \, , \, \frac{\ux_k}{1+bt} \, ; \, \frac{\ux_k'}{1+bt} \, \Big) \, 
	\nonumber\\
	&=&0
\eeqn
if $d=2$. This proves Theorem \ref{thm-pseudoconf-gamma-1}.
\qed

\newpage

\appendix

\section{The Klainerman-Machedon spacetime bounds} 
\label{sect-boundsfreeGP-1}

We present the Klainerman-Machedon spacetime bounds in dimensions $d\geq2$ in the form required for this paper,
with $\alpha\in\alphaset(d,p)$; see (\ref{eq-alphaset-def-1}). 
In the regime $\alpha>\frac d2-\frac{1}{2(p-1)}$, we present a simple 
argument to prove the result. 
In the endpoint case  $(d,p)=(3,2)$ and $\alpha=1$,
we invoke a result of \cite{klma}.
 
\begin{proposition}
\label{prp-spacetime-bd-1}
Let $p=2,4$ account for the cubic and quintic GP hierarchy, respectively, 
and assume that $\alpha\in\alphaset(d,p)$.
Let $\gamma^{(k+\frac p2)}$ be the solution of 
\eqn
	i\partial_t\gamma^{(k+\frac p2)}(t,\ux_{k+\frac p2};\ux_{k+\frac p2}')
	\, + \, (\Delta_{\ux_{k+\frac p2}}-\Delta_{\ux_{k+\frac p2}'})
	\gamma^{(k+\frac p2)}(t,\ux_{k+\frac p2};\ux_{k+\frac p2}') \, = \, 0
\eeqn
with initial condition
\eqn
	\gamma^{(k+\frac p2)}(0,\,\cdot\,) \, = \, \gamma_0^{(k+\frac p2)}\in\cH^\alpha \,.
\eeqn 
Then, there exists a constant $C$ such that 
\eqn
	\lefteqn{
	\Big\| \, S^{(k,\alpha)}B_{j;k+1,\dots,k+\frac p2}\gamma^{(k+\frac p2)} \, 
	\Big\|_{L^2 (\R\times\R^{dk}\times\R^{dk})}
	}
	\nonumber\\
	&&\quad\quad\quad\quad
	\, \leq \, C \, 
	\Big\| \,  S^{(k+\frac p2,\alpha)} \gamma_0^{(k+\frac p2)} \, 
	\Big\|_{L^2 (\R^{d(k+\frac p2)}\times\R^{d(k+\frac p2)})}  
\eeqn 
holds.
\end{proposition}

\prf
For notational convenience, we discuss the proof for the quintic GP hierarchy where $p=4$.

Let $(\tau,\uu_k,\uu_k')$, $\uq:=(q_1,q_2)$, and $\uq':=(q_1',q_2')$
denote the Fourier conjugate variables corresponding to 
$(t,\ux_k,\ux_k')$, $(x_{k+1},x_{k+2})$, and $(x_{k+1}',x_{k+2}')$, respectively.

Without any loss of generality, we may assume that $j=1$ in $B_{j;k+1,k+2}$. Then, 
abbreviating
\eqn
	\delta(\cdots) \, := \, \delta( \, \tau + (u_1+q_1+q_2-q_1'-q_2')^2  
	+ \sum_{j=2}^k u_j^2
	+ |\uq|^2 
	- |\uu_k'|^2 - |\uq'|^2   \, )
\eeqn
we find
\eqn
	\lefteqn{
	\Big\| \, S^{(k,\alpha)}B_{1;k+1,k+2}\gamma^{(k+2)} \, 
	\Big\|_{L^2 (\R\times\R^{d(k+2)}\times\R^{d(k+2)})}^2
	}
	\nonumber\\
	& = &
	\int_{\R}d\tau \int d\uu_k d\uu_k' \prod_{j=1}^k\bra u_j\ket^{2\alpha} \bra u_j'\ket^{2\alpha}
	\\
	&&\quad 
	\Big( \, \int d\uq d\uq' \, 
	\delta(\cdots)
	\widehat\gamma^{(k+2)}(\tau,u_1+q_1+q_2-q_1'-q_2',u_2,\dots,u_k,\uq;\uu_k',\uq')\, \Big)^2 \,.
	\nonumber
\eeqn
Using the Schwarz estimate, this is bounded by
\eqn
	&\leq&\int_{\R}d\tau \int d\uu_k d\uu_k' \, I_\alpha(\tau,\uu_k,\uu_k')
	\int d\uq d\uq' \,  
	\delta(\cdots)
	\nonumber\\
	&&
	\bra u_1+q_1+q_2-q_1'-q_2' \ket^{2\alpha} \bra q_1\ket^{2\alpha} \bra q_2\ket^{2\alpha}
	\bra q_1'\ket^{2\alpha} \bra q_2' \ket^{2\alpha}
	\prod_{j=2}^k\bra u_j\ket^{2\alpha} \prod_{j'=1}^k\bra u_{j'}'\ket^{2\alpha}
	\nonumber\\
	&&\quad\quad\quad\quad
	\Big| \, \widehat\gamma^{(k+2)}(\tau,u_1+q_1+q_2-q_1'-q_2',u_2,\dots,u_k,\uq;\uu_k',\uq') \, \Big|^2
\eeqn
where
\eqn\label{eq-Ialpha-def-1}
	\lefteqn{
	I_\alpha(\tau,\uu_k,\uu_k') 
	}
	\\
	&&
	\, := \,
	\int d\uq \, d\uq' \, 
	\frac{
	\delta(\cdots) \, \bra u_1\ket^{2\alpha}}
	{\bra u_1+q_1+q_2-q_1'-q_2'\ket^{2\alpha} \bra q_1\ket^{2\alpha} \bra q_2\ket^{2\alpha}
	\bra q_1'\ket^{2\alpha} \bra q_2'\ket^{2\alpha}} \,.
	\nonumber
\eeqn
Similarly as in \cite{klma,kiscst}, we observe that
\begin{equation}\label{eq-u1-aux-bd-1}
	 \bra u_1\ket^{2\alpha} 
	 \, \leq \, 
	 C\Big[ \, \bra u_1+q_1+q_2-q_1'-q_2'\ket^{2\alpha} 
	 + \bra q_1\ket^{2\alpha}
	 + \bra q_2\ket^{2\alpha}
	 + \bra q_1'\ket^{2\alpha} + \bra q_2'\ket^{2\alpha} \, \Big]
	 \,,
\end{equation} 
so that
\eqn
	I_\alpha(\tau,\uu_k,\uu_k') \, \leq \, \sum_{\ell=1}^5 J_\ell
\eeqn
where $J_\ell$ is obtained from bounding the numerator of
(\ref{eq-Ialpha-def-1}) using  (\ref{eq-u1-aux-bd-1}), 
and from canceling the $\ell$-th term on the rhs 
of (\ref{eq-u1-aux-bd-1}) with the corresponding term
in the denominator of (\ref{eq-Ialpha-def-1}).
Thus, for instance,
\eqn
	J_1 \, < \, \int d\uq \, d\uq' \, 
	\frac{
	\delta(\cdots)}
	{\bra q_1\ket^{2\alpha} \bra q_2\ket^{2\alpha}
	\bra q_1'\ket^{2\alpha} \bra q_2'\ket^{2\alpha}} \,,
\eeqn
and each of the terms $J_\ell$ with $\ell=2,\dots,5$ can be brought into
a similar form by appropriately translating one of the momenta $q_i$, $q_j'$.

Further following \cite{klma,kiscst}, we observe that the argument of the delta distribution
equals
\eqn
	\tau + (u_1+q_1+q_2-q_1')^2  
	+ \sum_{j=2}^k u_j^2
	+ |\uq|^2 
	- |\uu_k'|^2 - (q_1')^2 
	- 2(u_1+q_1+q_2-q_1')\cdot q_2' \,,
	\nonumber 
\eeqn
and we integrate out the delta distribution using the component of $q_2'$ parallel
to $(u_1+q_1+q_2-q_1')$. 
This leads to the bound
\eqn\label{eq-J1-bound-aux-1}
	J_1 
	& < &  
	C_\alpha C \int d\uq d q_1' \, 
	\frac{ 1}
	{| u_1+q_1+q_2-q_1'|
	\bra q_1\ket^{2\alpha} \bra q_2\ket^{2\alpha}
	\bra q_1'\ket^{2\alpha} } 
\eeqn
where
\eqn
	C_\alpha \, := \, \int_{\R} \frac{d\zeta}{\bra \zeta \ket^{2\alpha}} \,.
\eeqn 
Clearly, $C_\alpha$ is finite for any $\alpha>\frac12$.
Moreover, it is clear that $J_1$ is monotonically decreasing in $\alpha$.

For the cubic GP hierarchy, the above arguments lead to the condition that
instead of (\ref{eq-J1-bound-aux-1}),
the integral
\eqn\label{eq-J1-bound-aux-2} 
	\int  d q_1 \, 
	\frac{ 1}
	{| u_1+q_1 |
	\bra q_1\ket^{2\alpha}  } 
\eeqn
must be bounded.

\noindent\underline{Proof for $\alpha>\frac d2-\frac{1}{2(p-1)}$.}
\\
\\ 
We first consider the case $p=4$ corresponding to the quintic GP hierarchy, and argue as follows.
To bound  (\ref{eq-J1-bound-aux-1}), we pick a spherically symmetric  function 
$h\geq0$ with rapid decay away from the unit ball in $\R^d$, such that
$h^\vee(x)\geq0$ decays rapidly outside of the unit ball in $\R^d$, and
\eqn\label{eq-convuppbd-1}
	\frac{1}{\bra q \ket^{2\alpha}} \, < \, 
	  \, h*\Big( \, \chi_{B_1} \, + \, \frac{\chi_{B_1}^c}{| \, \cdot \, |^{2\alpha}} \, \Big)(q)   \,.
\eeqn 
(for example, $h(u)= c_1 e^{-c_2 u^2}$, for suitable constants $c_1,c_2$),
where $\chi_{B_1} +\chi_{B_1}^c=1$ is a smooth partition of unity with $\chi_{B_1}$
supported on the unit ball, with  $\chi_{B_1}(u)=1$ for $|u|\leq\frac12$, and  
$\chi_{B_1}(u)=0$ for $|u|>\frac12$.
Clearly, $h* \chi_{B_1}$ and $h*\frac{\chi_{B_1}^c}{| \, \cdot \, |^{2\alpha}}$
are both in $L^\infty$, for any $\alpha>0$. 

Then, assuming that $\alpha<\frac d2$, inserting this into (\ref{eq-J1-bound-aux-1}), 
the most singular part is given by
\eqn\label{eq-J1-convFTprod-bd-1}
	\lefteqn{ 
	C_\alpha C \Bra \,
	\Big(\frac{1}{| \, \cdot \, |}*(h*\frac{\chi_{B_1}^c}{| \, \cdot \, |^{2\alpha}}) \, \Big) *
	(h*\frac{\chi_{B_1}^c}{| \, \cdot \, |^{2\alpha}}) \, , \, (h*\frac{\chi_{B_1}^c}{| \, \cdot \, |^{2\alpha}}) \,
	\Ket_{L^2(\R^d)}
	}
	\nonumber\\
	&=& C_\alpha C \int dx \, \Big( \, \frac{\chi_{B_1}^c}{| \, \cdot \, |} \, \Big)^\vee(x)
	\, \Big( \, (h*\frac{\chi_{B_1}^c}{| \, \cdot \, |^{2\alpha}})^\vee(x) \, \Big)^3
	\nonumber\\
	&=& C_\alpha C' \int dx \,  \frac{1}{| \, x \, |^{d-1}}  (h^\vee(x))^3
	\, \Big( \, \Big( (\chi_{B_1}^c)^\vee*\frac{1}{| \, \cdot \, |^{d-2\alpha}} \Big) (x) \, \Big)^3
	\nonumber\\
	&<& C_\alpha C' \int dx \,  \frac{1}{| \, x \, |^{d-1}}  (h^\vee(x))^3
	\, \Big( \, \frac{1}{| \, x \, |^{d-2\alpha}} \, \Big)^3 \,.
\eeqn
For sufficiently large $C'$, this is an upper bound on all of the remaining terms
that are obtained from substituting the bound (\ref{eq-convuppbd-1})  into (\ref{eq-J1-bound-aux-1}).
We have here used that $(\chi_{B_1}^c)^\vee=1^\vee-\chi_{B_1}^\vee=\delta-\chi_{B_1}^\vee$, 
so that $|((\chi_{B_1}^c)^\vee*\frac{1}{| \, \cdot \, |^{d-2\alpha}})(x)| \leq C \, \frac{1}{|x|^{d-2\alpha}}$
holds for $\alpha<\frac d2$.  

We conclude that (\ref{eq-J1-convFTprod-bd-1}) is finite provided that the singularity at $x=0$ is integrable,
since $h^\vee(x)$ falls off rapidly as $|x|\rightarrow\infty$.
In dimension $d$, this is the case if the exponents in the denominator satisfy
\eqn
	d-1 \, + \, 3d \, - \, 6\alpha \, < \, d \,,
\eeqn
such that
\eqn\label{eq-alpha-lowbd-1}
	\alpha \, > \, \frac d2-\frac16 \,. 
\eeqn
This proves the claim for the quintic GP hierarchy, i.e., for $p=4$.
In order to prove the lower bound (\ref{eq-alpha-lowbd-1}) on $\alpha$, we 
have assumed that $\alpha<\frac d2$, which is consistent with it.
Now, since  $J_1$ is monotonically decreasing in $\alpha$,  we arrive at the asserted result.

For the cubic GP hierarchy, the same considerations lead to the condition that
$(\ref{eq-J1-bound-aux-1}) \,  < \, \infty$ if
$\alpha \, > \, \frac d2-\frac12$. 
For a general $p$-GP hierarchy, one obtains the condition $\alpha>\frac d2-\frac{1}{2(p-1)}$.

\noindent\underline{The case $\alpha=1$ for the cubic GP hierarchy in $d=3$.}
\\
\\
In the situation $d=3$ and $p=2$ of the cubic GP hierarchy in 3 dimensions,
we have the endpoint case  $\frac d2-\frac1{2(p-1)}=1$.
Klainerman and Machedon have proven in \cite{klma} that 
(\ref{eq-Ialpha-def-1}) is bounded in this case. 
\endprf

Next, we prove the iterated spacetime estimates for the cubic GP hierarchy used
in Section \ref{sec-locwp-1}, which involve the "boardgame estimates" of
\cite{klma}, which are motivated by the Feynman graph techniques in \cite{esy1,esy2,ey}. 
The corresponding results for the quintic GP hierarchy are 
obtained in an analogous manner, and we refer to \cite{chpa} for details.

\begin{proposition}
\label{prp-spacetime-bd-2}
Assume $\alpha$ as in Proposition \ref{prp-spacetime-bd-1},
for the cubic GP hierarchy ($p=2$). For brevity, let
\eqn
	H^\alpha_k \, := \, H^\alpha(\R^{dk}\times\R^{dk}) \,.
\eeqn
Then, for $k\geq1$ and $j\leq k$, and $t\in I=[0,T]$,
\eqn
	\| \, B_{k+1}\duh_j(\Gamma_0)^{(k+1)}(t) \, \|_{L^1_{t\in I}H^\alpha_k} 
	\, < \,  k C^k (c T)^{\frac {j+1}2} \|\gamma^{(k+j+1)}_0\|_{H^\alpha_{k+j+1}} \,.
\eeqn
Moreover,
\eqn\label{eq-Boardgame-rem-est-1}
	\| \,  B_{k+1}\duh_{k}(\opB\Gamma)^{(k+1)} \, \|_{L^1_{t\in I}H^\alpha_k} 
	\, < \,  k C^k (c T)^{\frac k2} \| \, (\opB\gamma)^{(2k)} \, \|_{L^1_{t \in I}H^\alpha_{2k}} \,,
\eeqn
where $\duh_j( \, \cdot \,)$ is defined in (\ref{eq-Duh-j-def-1}).
\end{proposition}

\prf
Let $I=[0,T]$.
Using an argument presented as a ``board game", 
it is proven in \cite{klma} that the following holds.

Let ${\mathcal E}_{j,k+1}$ denote the space of sequences $\umu=(\mus(1),\dots,\mus(j))$
where $\mu(i)\in\{1,\dots,k+i\}$, where for every $i\in\{1,\dots,j\}$,
one has $\mu(i)\geq\mu(i')$ for all $i'>i$. The elements of ${\mathcal E}_{j,k+1}$ 
parametrize $(k+j)\times j$ matrices in so-called ``special upper echelon form"
(see \cite{klma} for definitions).
The cardinality of this set satisfies $|{\mathcal E}_{j,k+1}|\leq C^{j+k}$.

For every $\umu\in{\mathcal E}_{j,k+1}$, one associates the term
\eqn
	\lefteqn{
	( \, \duh_j(\Gamma_0)^{(k+1)}(t) \, )_{\umu}
	}
	\nonumber\\
	&:=&\int_{D} dt_1 \cdots  dt_{j}
	e^{i(t-t_1)\Delta_\pm^{(k+1)}}B_{\mu(1),k+2}e^{i(t_1-t_2)\Delta_\pm^{(k+2)}}\cdots
	\nonumber\\
	&&\quad
	\cdots
	 e^{i(t_{j-1}-t_{j})\Delta_\pm^{(k+j)}} 
        B_{\mus(j),k+j+1}e^{i t_j \Delta_\pm^{(k+j+1)}}  {\gamma}^{(k+j+1)}_0 \, 
\eeqn 
for a measurable subset $D\subset[0,t]^j$.
Then, it is proven in \cite{klma} that
\eqn\label{eq-klma-boardgame-dec-1}
	\lefteqn{
	\| \, B_{k+1}\duh_j(\Gamma_0)^{(k+1)}(t) \, \|_{L^2_{t\in I}H^\alpha} 
	}
	\nonumber\\
	&\leq & \sum_{\umu\in {\mathcal E}_{j,k+1}} 
	\| \, B_{k+1} ( \, \duh_j(\Gamma_0)^{(k+1)}(t) \, )_{\umu} \, \|_{L^2_{t\in I}H^\alpha} \,.
\eeqn 
For the proof of (\ref{eq-klma-boardgame-dec-1}) in the case of the cubic GP hierarchy,
we refer to \cite{klma}. For the case of the quintic GP hierarchy, we refer to 
\cite{chpa}.

We have, under the given assumptions on $\alpha$,
that for $I=[0,T]$ and $D\subset I^j$,
\eqn
	\lefteqn{
	\| \, B_{k+1} ( \, \duh_j(\Gamma_0)^{(k+1)}(t) \, )_{\underline{\mu}_s} \, \|_{L^1_{t\in I}H^\alpha} 
	}
	\nonumber\\
	&\leq& \sum_{\ell=1}^k
	\Big\| \, \int_{D} dt_1 \cdots  dt_{j} B_{\ell,k+1}e^{i(t-t_1) \Delta_\pm^{(k+1)}}
	B_{\mus(1),k+2}e^{i(t_1-t_2)\Delta_\pm^{(k+2)}}\cdots
	\nonumber\\
	&&\quad
	\cdots
	e^{i(t_{j-1}-t_{j})\Delta_\pm^{(k+j)}} 
        B_{\mus(j),k+j+1} e^{i t_j \Delta_\pm^{(k+j+1)}} 
        {\gamma}^{(k+j+1)}_0\, \Big\|_{L^1_{t\in I}H^\alpha_k}
	\nonumber\\
	&\leq& k
	\, \int_{I} dt \int_D dt_1 \cdots  dt_{j} 
	\Big\| \, B_{\ell,k+1}e^{i(t - t_1)\Delta_\pm^{(k+1)}}B_{\mus(1),k+2}e^{i(t_1-t_2)\Delta_\pm^{(k+2)}}
	\cdots
	\nonumber\\
	&&\quad
	\cdots
	e^{i(t_{j-1}-t_{j})\Delta_\pm^{(k+j)}} 
        B_{\mus(j),k+j+1}
         e^{i t_j \Delta_\pm^{(k+j+1)}} {\gamma}^{(k+j+1)}_0\, \Big\|_{ H^\alpha_k}
	\nonumber\\
	&\leq& k
	\, \int_{I^{j+1}} dt   dt_1 \cdots  dt_{j} 
	\Big\| \, B_{\ell,k+1}e^{i(t - t_1)\Delta_\pm^{(k+1)}}B_{\mus(1),k+2}e^{i(t_1-t_2)\Delta_\pm^{(k+2)}}
	\cdots
	\nonumber\\
	&&\quad
	\cdots
	e^{i(t_{j-1}-t_{j})\Delta_\pm^{(k+j)}} 
        B_{\mus(j),k+j+1} 
         e^{i t_j \Delta_\pm^{(k+j+1)}}{\gamma}^{(k+j+1)}_0\, \Big\|_{ H^\alpha_k} \,.
         \label{eq-BGamma-Duhj-aux-1}
\eeqn
By Cauchy-Schwarz with respect to the integral in $t$, this is bounded by
\eqn
	&\leq& k T^{\frac12}
	\, \int_{I^{j}}    dt_1 \cdots  dt_{j} 
	\Big\| \, B_{\ell,k+1}e^{i(t - t_1)\Delta_\pm^{(k+1)}}B_{\mus(1),k+2}e^{i(t_1-t_2)\Delta_\pm^{(k+2)}}
	\cdots
	\nonumber\\
	&&\quad
	\cdots
	e^{i(t_{j-1}-t_{j})\Delta_\pm^{(k+j)}} 
        B_{\mus(j),k+j+1}
         e^{i t_j\Delta_\pm^{(k+j+1)}} {\gamma}^{(k+j+1)}_0\, \Big\|_{ L^2_t(I)H^\alpha_k}
	\nonumber\\
	&\leq& k T^{\frac12}
	\, \int_{I^{j}}    dt_1 \cdots  dt_{j} 
	\Big\| \, B_{\ell,k+1}e^{i(t - t_1)\Delta_\pm^{(k+1)}}B_{\mus(1),k+2}e^{i(t_1-t_2)\Delta_\pm^{(k+2)}}
	\cdots
	\nonumber\\
	&&\quad
	\cdots
	e^{i(t_{j-1}-t_{j})\Delta_\pm^{(k+j)}} 
        B_{\mus(j),k+j+1}
         e^{i t_j \Delta_\pm^{(k+j+1)}} {\gamma}^{(k+j+1)}_0\, \Big\|_{ L^2_t(\R)H^\alpha_k} \,. \quad
\eeqn
Using Proposition \ref{prp-spacetime-bd-1} and unitarity of $e^{-i t_1\Delta_\pm^{(k+1)}}$, this is bounded by
\eqn	
	&\leq& 	k (cT)^{\frac12} \, \int_{I^{j}} dt_1 \cdots  dt_{j}
	\Big\| \,B_{\mus(1),k+2}e^{i(t_1-t_2)\Delta_\pm^{(k+2)}} \cdots
	\nonumber\\
        &&\quad
	\cdots
	e^{i(t_{j-1}-t_{j})\Delta_\pm^{(k+j)}} 
        B_{\mus(j),k+j+1}
         e^{i t_j \Delta_\pm^{(k+j+1)}} {\gamma}^{(k+j+1)}_0\, \Big\|_{ H^\alpha_{k+1}}
	\label{ap-usingKM1}\\
	&=& 	k  \, (c T)^{\frac12}\, \int_{I^{j-1}} dt_2 \cdots  dt_{j} 	
	\Big\| \, B_{\mus(1),k+2} e^{i(t_1-t_2)\Delta_\pm^{(k+2)}}  \cdots
	\nonumber\\
	&& \quad 
        \cdots
	e^{i(t_{j-1}-t_{j})\Delta_\pm^{(k+j)}} 
        B_{\mus(j),k+j+1}
         e^{i t_j \Delta_\pm^{(k+j+1)}} 
         {\gamma}^{(k+j+1)}_0 \, \Big\|_{L^1_{t_1}(I)H^\alpha_{k+1}} \,. \quad\quad
	\label{ap-Hol}
\eeqn
Iterating the same steps as above, we find, after $j$ steps,
\eqn
        & \leq & \cdots\cdots\cdots
        \nonumber \\
	&\leq& k  (c T)^{\frac{j}{2}} \int_{I} dt_{j} 
        \Big\|B_{\mus(j-1),k+j} e^{i(t_{j-1}-t_{j})\Delta_\pm^{(k+j)}} 
        \label{ap-KMiterated}\\
        &&\quad\quad\quad\quad\quad\quad\quad
        B_{\mus(j),k+j+1}  e^{i t_j \Delta_\pm^{(k+j+1)}}
        {\gamma}^{(k+j+1)}_0 \Big\|_{L^2_{t_{j-1}\in I}H^\alpha_{k+j-1}}
        \nonumber \\
        &\leq& k  (c T)^{\frac{j}{2}} \int_{I} dt_{j} 
        \Big\| B_{\mus(j),k+j+1} 
        e^{i t_{j} \Delta_\pm^{(k+j+1)}} 
        {\gamma}^{(k+j+1)}_0 \Big\|_{H^\alpha_{k+j}}
        \nonumber \\ 
        & \leq & k  (c T)^{\frac{j+1}{2}} 
        \Big\| {\gamma}^{(k+j+1)}_0 \Big\|_{H^\alpha_{k+j+1}} \,.  
        \label{A2final} 
\eeqn
In the last step, we used Cauchy-Schwarz in $t_j$, and Proposition \ref{prp-spacetime-bd-1}.

Then, estimating by $C^{j+k}$ the number of terms in the sum  over $\umu\in {\mathcal E}_{j,k+1}$,
\eqn\label{eq-BGamma-Duhj-combin-bd-1}
	\| \, B_{k+1}\duh_j(\Gamma_0)^{(k+1)}(t) \, \|_{L^1_{t\in I}H^\alpha} 
	\, \leq \, k C^k (c T)^{\frac {j+1}2} \|\gamma^{(k+j+1)}_0\|_{H^\alpha}  \,,
\eeqn
as claimed.

In the same manner, we prove (\ref{eq-Boardgame-rem-est-1}). In this case, we have
\eqn
	\lefteqn{
	\| \, B_{k+1} ( \, \duh_k(\opB\Gamma )^{(k+1)}(t) \, )_{\underline{\mu}_s} \, \|_{L^1_{t\in I}H^\alpha_k} 
	}
	\nonumber\\
	&\leq& \sum_{\ell=1}^k
	\Big\| \, \int_{D} dt_1 \cdots  dt_{k} B_{\ell,k+1}e^{i(t-t_1) \Delta_\pm^{(k+1)}}
	B_{\mus(1),k+2}e^{i(t_1-t_2)\Delta_\pm^{(k+2)}}\cdots
	\nonumber\\
	&&\quad
	\cdots
	e^{i t_{k} \Delta_\pm^{(2k)}} 
    B_{ 2k+ 1}   {\gamma}^{(2k+ 1)}(t_k)\, \Big\|_{L^1_{t\in I}H^\alpha_k}
    \nonumber\\
	&\leq& k \, \int_{I}dt
	 \int_{D} dt_1 \cdots  dt_{k} \Big\| \, B_{\ell,k+1}e^{i(t-t_1) \Delta_\pm^{(k+1)}}
	B_{\mus(1),k+2}e^{i(t_1-t_2)\Delta_\pm^{(k+2)}}\cdots
	\nonumber\\
	&&\quad
	\cdots
	e^{i t_{k} \Delta_\pm^{(2k)}} 
        B_{ 2k+ 1}  
        {\gamma}^{(2k+ 1)}(t_k)\, \Big\|_{ H^\alpha_k}
\eeqn
Applying the same  arguments as above between (\ref{eq-BGamma-Duhj-aux-1})
and  (\ref{ap-KMiterated}), with $j=k$, one finds the upper bound
\eqn
        & \leq & \cdots\cdots\cdots
        \nonumber \\
	&\leq& k  (c T)^{\frac{k}{2}} \int_{I} dt_{k} 
        \Big\|B_{\mus(k-1),2k } e^{i t_{k} \Delta_\pm^{(2k )}} 
        \label{ap-KMiterated-BGamma}\\
        &&\quad\quad\quad\quad\quad\quad\quad\quad\quad
        B_{ 2k+ 1}  
        {\gamma}^{(2k+ 1)}(t_k) \Big\|_{L^2_{t_{k-1}\in I}H^\alpha_{2k -1}}
        \nonumber \\
        &\leq& k  (c T)^{\frac{k}{2}} \int_{I} dt_{k} 
        \Big\| B_{ 2k+ 1}  
        {\gamma}^{(2k+ 1)}(t_k) \Big\|_{H^\alpha_{2k}}
        \nonumber \\ 
        & = & k  (c T)^{\frac{k}{2}} 
        \Big\| B_{ 2k+ 1}  
        {\gamma}^{(2k+ 1)}(t_k) \Big\|_{L_{t\in I}^1H^\alpha_{2k }}
        \nonumber \\ 
        & = & k  (c T)^{\frac{k}{2}} 
        \Big\| (\opB\Gamma)^{(2k)}(t_k) \Big\|_{L_{t\in I}^1H^\alpha_{2k }} \,.  
        \label{A2final-BGamma} 
\eeqn
Invoking the argument used to establish (\ref{eq-BGamma-Duhj-combin-bd-1}), we
arrive at  (\ref{eq-Boardgame-rem-est-1}).

For more details, we refer to \cite{chpa}.
\endprf

\subsection*{Acknowledgements}
We thank S. Klainerman and N. Tzirakis for inspiring discussions. 
We are deeply indebted to B. Schlein for extremely useful comments that
helped us to significantly improve a previous version.
The work of T.C. was supported by NSF grant DMS-0704031.
The work of N.P. was supported NSF grant number DMS 0758247 
and an Alfred P. Sloan Research Fellowship.

\end{document}